\documentclass[fleqn,usenatbib,useAMS]{mnras}

\usepackage{graphicx}	
\usepackage{amsmath}	
\usepackage{amssymb}	
\usepackage{multicol}        
\usepackage{bm}		
\usepackage{pdflscape}	

\graphicspath{{figures/}}
\newcommand{\M}[1]{M_{\rm #1}}
\newcommand{\ergs}{erg\,s$^{-1}$}

\usepackage[T1]{fontenc}
\usepackage{ae,aecompl}

\usepackage{newtxtext,newtxmath}

\title[AGN-Merger Connection]{The Connection between Mergers and AGN Activity in Simulated and Observed Massive Galaxies}

\author[Sharma et al.]{
Ray S. Sharma$^{1}$,
Ena Choi$^{2}$ \thanks{E-mail: enachoi@uos.ac.kr},
Rachel S. Somerville$^{3}$,
Gregory F. Snyder$^{4}$,
Hannah Jhee$^{2}$,
\newauthor{
Dale D. Kocevski$^{5}$,
Michaela Hirschmann$^{6,7}$,
Benjamin P. Moster$^{8,9}$,
Thorsten Naab$^{9}$}
\newauthor{
Desika Narayanan$^{10,11,12}$,
Jeremiah P. Ostriker$^{13}$, David J. Rosario$^{14}$}
\\
$^{1}$Department of Physics and Astronomy, Rutgers, The State University of New Jersey, 136 Frelinghuysen Road, Piscataway, NJ 08854, USA\\
$^{2}$Department of Physics, University of Seoul, 163 Seoulsiripdaero, Dongdaemun-gu, Seoul 02504, Republic of Korea\\
$^{3}$Center for Computational Astrophysics, Flatiron Institute, 162 5th Ave., 
    New York, NY 10010, USA \\
$^{4}$Space Telescope Science Institute, 3700 San Martin Dr, Baltimore, MD 21218, USA\\
$^{5}$Department of Physics and Astronomy, Colby College, Waterville, ME 04961, USA\\
$^{6}$Institut de Physique, Laboratoire d'astrophysique, \'Ecole Polytechnique F\'ed\'erale de Lausanne (EPFL), CH-1290 Versoix, Switzerland\\
$^{7}$DARK, Niels Bohr Institute, University of Copenhagen, Lyngbyvej 2, DK-2100 Copenhagen, Denmark\\
$^{8}$Universit{\"a}ts-Sternwarte, Ludwig-Maximilians-Universit{\"a}t M{\"u}nchen, Scheinerstr. 1, 81679 M{\"u}nchen, Germany\\
$^{9}$Max-Planck-Institut f{\"u}r Astrophysik, Karl-Schwarzschild Stra{\ss}e 1, 85748 Garching, Germany\\
$^{10}$Department of Astronomy, University of Florida, 211 Bryant Space Sciences centre, Gainesville, FL 32611, USA\\
$^{11}$University of Florida Informatics Institute, 432 Newell Drive, CISE Bldg E251, Gainesville, FL 32611, USA\\
$^{12}$Cosmic Dawn Center at the Niels Bohr Institute, University of Copenhagen and RFU-Space, Technical University of Denmark\\
$^{13}$Department of Astronomy, Columbia University, New York, NY 10027, USA\\
$^{14}$School of Mathematics, Statistics and Physics, Newcastle University, Newcastle upon Tyne, NE1 7RU, UK
}

\date{Last updated 2023 November 17; in original form 2023 April 10}

\pubyear{2023}

\begin{document}
\label{firstpage}
\pagerange{\pageref{firstpage}--\pageref{lastpage}}
\maketitle

\begin{abstract}
    We analyze a suite of $29$ high-resolution zoom-in cosmological hydrodynamic simulations of massive galaxies with stellar masses $M_{\rm star} > 10^{10.9} M_\odot$, with the goal of better understanding merger activity among AGN, AGN activity in merging systems, SMBH growth during mergers, and the role of gas content in triggering AGN. Using the radiative transfer code \textsc{Powderday}, we generate HST-WFC3 F160W mock observations of central galaxies at redshift $0.5 < z < 3$; convolve each image with a CANDELS-like PSF; stitch each image into a real CANDELS image; and identify mergers within the synthetic images using commonly adopted non-parametric statistics. We study the connection between mergers and AGN activity in both the simulations and synthetic images and find reasonable agreement with observations from CANDELS. We find that AGN activity is not primarily driven by major mergers (stellar mass ratio $>$1:4) except in a select few cases of gas-rich mergers at low redshifts $(0.5 < z < 0.9)$. We also find that major mergers do not significantly grow the central SMBHs, indicating major mergers do not sustain long-term accretion. Moreover, the most luminous AGN in our simulations $(L_{\rm bol} > 10^{45}$ erg s$^{-1})$ are no more likely than inactive galaxies $(L_{\rm bol} < 10^{43}$ erg s$^{-1})$ to be found in merging systems. We conclude that mergers are not the primary drivers of AGN activity in the simulated massive galaxies studied here.
\end{abstract}


\section{Introduction}

    Supermassive black holes (SMBHs) are thought to be ubiquitous within massive galaxies \citep{Magorrian1997,Kormendy2013}. However, it is still a mystery why gas accretion rates onto these SMBH are low to negligible in some galaxies and extremely high in others. In order to power the most luminous quasars, nearly the whole gas content of a massive galaxy must be funneled onto the black hole in a few hundred million years, or roughly a dynamical time \citep{Hopkins2008a}. The primary mechanism that triggers episodes of rapid accretion, which give rise to the observational signatures of active galactic nuclei (AGN), is still unclear. Galaxy interactions offer a compelling candidate for a number of reasons. Mergers and interactions can dissipate angular momentum and channel gas into galactic nuclei \citep{Barnes1991,Springel2005a,Gabor2016,Blumenthal2018}. This may simultaneously funnel gas onto the central SMBH and drive star formation in the bulge, helping to explain tight observed correlations between SMBH and galaxy bulge properties \citep{Gebhardt2000,Ferrarese2000,Marconi2003,Haring2004}.
    
    Furthermore, major mergers can transform rotation-supported disk-dominated galaxies into pressure-supported spheroid-dominated galaxies, while AGN feedback may efficiently heat and drive gas out of the galaxy, leading to the quenching of star formation. Thus, a link between mergers and AGN triggering could explain the observed correlation between spheroid-like morphologies and kinematics, and old stellar populations \citep[][and references therein]{Somerville2015,Kormendy2013}. It has been shown using empirical and semi-analytic models that, under the ansatz that mergers trigger AGN activity, the observed merger rate can account for the observed evolution of the AGN luminosity function \citep{Hopkins2006a,Hopkins2008a}, as well as the observed growth rate of quenched, spheroid dominated galaxies \citep{Hopkins2006a,Hopkins2008b,Somerville2008,Brennan2015}. 
    
    However, there are numerous other possible mechanisms for fueling accretion onto black holes and, thereby AGN activity. Alternative mechanisms that can funnel gas towards the centres of galaxies have been identified, including secular processes such as bar instabilities \citep{Kormendy2004,Hirschmann2012a} and violently unstable and/or clumpy disks \citep{Dekel2009,Bournaud2011,Mullaney2012a}. In addition, recycled gas from the evolution of stars in the galaxy's nucleus can fuel significant AGN activity \citep{Ciotti1997,Ciotti2007}. Cooling flows from the hot halo are another proposed fueling mechanism \citep{Nulsen2000}. Recent work using high-resolution simulations of massive ellipticals has highlighted the importance of cooling flows fueled by stellar evolution in generating AGN outbursts \citep{Gan2019,Pellegrini2020,Gan2020}.
    
    Observations do not clearly show the connection between AGN and galaxy mergers. Many observational studies, using a variety of techniques, find mergers to be correlated with AGN activity \citep[e.g.][]{Sanders1988,Koss2012,Satyapal2014}. Many studies have shown that AGN have a higher interaction fraction of mergers (or interaction) than non-AGN control samples in the local universe \citep[e.g.][]{Koss2010,Cotini2013} and intermediate redshift range \citep[e.g.][]{Bessiere2012}. Other studies showed that galaxies exhibiting merger features have higher AGN fractions than non-interacting control samples in the local universe \citep{Ellison2011,Weston2017,Ellison2019}, and intermediate redshift range \citep{Silverman2011,Lackner2014}. On the other hands, many observational studies find that mergers do not correlate with AGN activity \citep[][]{DeRobertis1998,Li2008,Schawinski2012}. Especially at high redshifts, the AGN-merger connection appears to be weak -- specifically, many observational studies have not found a difference in the merger fraction between AGN and mass-matched non-AGN samples \citep{Kocevski2012,Mechtley2016,Marian2019}. A study comparing X-ray selected AGN and inactive galaxies in the intermediate redshift range of $z<1$ found a similar trend \citep{Grogin2005,Gabor2009,Cisternas2011,Villforth2014a}. Even at low redshifts, many studies did not find differences in interaction and/or merger fractions between active and inactive galaxies \citep[e.g.][]{Reichard2009,Sabater2015}. Studies supporting an AGN-merger connection typically find the most luminous AGN ($L_{\rm bol} > 10^{45}\;{\rm erg/s}$) are preferentially found in disturbed systems \citep{Treister2012,Glikman2015,Goulding2018}. Further, AGN with significant obscuration are more likely than unobscured AGN to be found in galaxies showing evidence of interactions or merging \citep{Glikman2015,Fan2016,Weston2017,Goulding2018,Donley2018}.

    Simulations draw similar mixed conclusions on the AGN-merger connection. Studies using hydrodynamic simulations of binary mergers of idealized galaxies demonstrated clear triggering of gas inflows to the nucleus and resulting starburst and AGN activity \citep{Springel2005,DiMatteo2005,Hopkins2005,Choi2012}.
    \citet{Dubois2015} find that mergers are the dominant mechanism for AGN fueling within their zoom-in cosmological simulation of $10^{12} M_\odot$ halo at $z=2$. Once a galaxy grows above stellar mass $M_{\rm star} > 10^{9} M_\odot$ threshold, they found that galaxies can confine nuclear inflows from mergers and sustain BH accretion. \citet{McAlpine2020} find an over-abundance of luminous AGN in disturbed systems with $M_{\rm star} \sim 10^{10} M_\odot$ within the {\sc Eagle} cosmological hydrodynamic simulation \citep{Schaye2015,Crain2015}, with AGN activity peaking immediately after galaxy coalescence. However, they also find the majority of SMBH growth does not occur during mergers.
    
    Some simulations find that mergers are difficult to disentangle from smooth gas accretion, as these processes tend to occur together in cosmological simulations. \citet{Bellovary2013} trace the origins of gas particles accreted by SMBHs in cosmological zoom simulations and find SMBHs are not preferentially fueled by either smoothly-accreted or merger-accreted gas. Similarly, \citet{Menci2014} use semi-analytic models to investigate disk instabilities in AGN, finding that mergers and disk instabilities are viable candidates as the dominant fueling mechanism. 
    
    Finally, other simulations find mergers are not the primary mode by which AGN are triggered. \citet{Steinborn2018} find that merger fractions in the Magneticum Pathfinder cosmological simulation are three times higher within AGN hosts than inactive galaxies. However, they find merging systems make up less than $20\%$ of AGN hosts, hence mergers play a minor role in driving the overall AGN population. \citet{Hirschmann2012a} show that semi-analytic models where SMBH growth is driven by mergers can reproduce the AGN luminosity function at intermediate redshifts and luminosities, but an additional accretion mode via disc instabilities is required in order to fully reproduce the number densities of the most luminous AGN.

    There are a number of explanations for the confusing status of the question of whether mergers play a vital role in AGN activity and BH growth. First, selecting AGN is not a straightforward task. AGN emission spans a broad range of wavelengths, and the radiation can originate from many different phenomena, including the accretion disk, dust obscuring structures, and jets. Second, identifying galaxy mergers is also not a straightforward task. Mergers have been commonly identified by visual inspection \citep{Bridge2007,Jogee2009,Darg2010,Kartaltepe2015,Rosario2015}, close pairs \citep{Lackner2014,Mundy2017,Mantha2018}, non-parametric morphological diagnostics \citep{Conselice2003,Lotz2004,Pawlik2016,Nevin2019}, and statistical learning techniques \citep{Hocking2018,Goulding2018,Ackermann2018,Snyder2019}.
    Each method may be particularly sensitive or insensitive to certain merger stages or may be limited by the availability of data. Few previous studies have attempted to select both mergers and AGN \emph{in the same way} in an observational sample and a cosmological simulation in order to robustly determine whether the AGN-merger connection seen in current simulations is consistent with observations.
    
    In this work, we analyze a set of zoom-in cosmological simulations to better understand the connection between mergers and AGN activity. In particular, we aim to answer whether 1) AGN are more likely to be found in disturbed or merging galaxies than a control sample, and 2) mergers/disturbed galaxies are more likely to host AGN. Studying high-resolution galaxy simulations in a cosmological context allows us to better understand how AGN may be triggered by mergers, how SMBHs grow during periods of merger activity, and how gas content plays into SMBH fueling. To select mergers in the same way as observers do, we mock-observe the simulations and generate noise-added images similar to HST-WFC3 F160W images. By comparing both intrinsic and synthetic merger statistics, we hope to understand better the role of observational selection effects in this problem. We compute a standard suite of non-parametric morphological statistics on both the synthetic images and the real CANDELS images and use machine learning to identify selection criteria that more accurately and robustly identify true mergers in the simulations. 

    In Section \ref{Simulation}, we summarize the basic physical ingredients of the cosmological zoom-in simulations used in this study. In Section \ref{Methods}, we detail the pipeline for generating synthetic images, adding noise, and measuring morphological parameters. We also detail the training and testing of an automated classifier for identifying mergers from synthetic images. In Section \ref{Results}, we explore how merger statistics vary for AGN versus inactive galaxies. We also explore how mergers influence AGN activity, how they impact overall SMBH growth, and what role gas plays in the fueling of AGN.

\section{Simulation} \label{Simulation}

    We use the cosmological zoom-in simulations of massive galaxy formation presented in \citet{Choi2017}. Here we give a brief summary of the code basics, physics, and feedback models we use in the simulation, but we refer the readers to \citet{Choi2017} for a detailed description of the simulations.

    \subsection{Simulation properties}

        The SPHGal \citep{Hu2014} code, a modified version of the parallel smoothed particle hydrodynamics (SPH) code GADGET-3 \citep{Springel2005b}, was used to run the simulations. SPHGal includes many improvements to overcome the numerical fluid-mixing problems of classical SPH codes, such as a density-independent pressure-entropy SPH formulation and an improved artificial viscosity implementation \citep[see][for details]{Hu2014}. The initial conditions are drawn from \citet{Oser2010} with the cosmological parameters of WMAP3 \citep[][$h=0.72$, $\Omega_{\rm b}=0.044$, $\Omega_{\rm dm}=0.26$, $\Omega_{\Lambda}=0.74$, $\sigma_{\rm 8}=0.77$, and $n_{\rm s}=0.95$]{Spergel2007}. In this work, we use cosmological zoom-in hydrodynamic simulations of $29$ massive halos adopting the fiducial model of \citet{Choi2017}. The total masses of the central halos at $z=0$ are $1.4 \times 10^{12} \le M_{\rm vir}\,/\,M_{\odot} \le 2.3 \times 10^{13}$, and stellar masses are $8.2 \times 10^{10} \le M_{\rm star}\,/\,M_{\odot} \le 1.5 \times 10^{12} $. The baryonic (star and gas particle) mass resolution is $m_{\rm star,gas}=5.8 \times 10^{6} \thinspace { \rm M_{\odot}}$, and the dark matter particle resolution is $m_{\mathrm{dm}} = 3.4 \times 10^{7} \thinspace{ \rm M_{\odot}}$. The comoving gravitational softening lengths are $\epsilon_{\mathrm{gas,star}} = 0.556 {\rm\thinspace kpc} $ for the gas and star particles and $\epsilon_{\mathrm{halo}} = 1.236 {\rm\thinspace kpc}$ for the dark matter particles. Halos are identified on the fly within the zoom-in simulation using a friends-of-friends algorithm at regular time intervals.

        The simulations include chemical evolution, metal cooling, and diffusion. The chemical evolution of 11 different species, H, He, C, N, O, Ne, Mg, Si, S, Ca, and Fe, are traced explicitly for star and gas particles with the chemical yields adopted respectively from \citet{Iwamoto1999,Woosley1995,Karakas2010} for Type I, Type II SNe and AGB stars. In addition, the simulations include a redshift-dependent UV/X-ray cosmic microwave background from \citet{Haardt2001a}, and the cooling rate from \citet{Wiersma2009} for optically thin gas in ionization equilibrium. Finally, the metal diffusion model is adopted from \cite{Aumer2013}, and allows the metal-enriched gas particles to mix their metals with neighbouring gas particles via turbulent diffusion.

    \subsection{Star Formation and Stellar Feedback}
        
        The star formation rate is calculated as $d \rho_{\rm star} /dt = \eta \rho_{\rm gas} / t_{\rm dyn}$ where $\rho_{\rm star}$,  $\rho_{\rm gas}$ and $t_{\rm dyn}$ are the stellar and gas densities, and local dynamical time for gas particle respectively, with the star formation efficiency set to $\eta = 0.025$. The simulation stochastically forms star particles in high-density regions where the gas density exceeds a density threshold, $n_{\rm th} \equiv n_0 \left( T_{\rm gas}/ T_0 \right)^3 \left( M_0 / M_{\rm gas}\right)^2 $. The critical threshold density and temperature are $n_0 = 2.0$ cm$^{-3}$ and $T_0 = 12000\,$K and $M_0$ is the gas particle mass at the fiducial resolution. 

        The simulations include the stellar feedback model of \citet{Nunez2017}, which consists of the winds from young massive stars, UV heating within Str{\"o}mgren spheres of young stars, the three-phase supernova feedback from type I and type II SNe, and metals and outflows from low-mass AGB stars.

        Before the moment of the SN explosion, the young stellar particles inject the same amount of mass and momentum as those of type II SN explosions evenly spread in time to the surrounding ISM via winds from young massive stars. Their ionizing radiation also gradually heats the neighbouring gas particles to $T=10^4$~K within a Str\"{o}mgren radius \citep{Stromgren1939}.

        The SNe model assumes that a single SN event ejects mass in an outflow with a velocity $v_{\rm out,SN}=4,500 {\rm\thinspace km~s^{-1}}$ to the surrounding ISM when star particles explode as SNe. Each neighbouring gas particle is affected by one of the three successive phases of SN depending on the physical distance from the SN particle: (i)~momentum-conserving free expansion phase, (ii)~energy-conserving Sedov-Taylor phase where SN energy is transferred with 30\% as kinetic and 70\% as thermal, and (iii)~the snowplow phase where radiative cooling becomes dominant.
        Then, both kinetic and thermal energy modes dissipate with distance from the SN in the pressure-driven snowplow phase of SN remnants. 

        Even after a SN event, old stellar particles keep distributing energy, momentum, and metals via AGB winds. The outflowing wind velocity of AGB stars is assumed to be $v_{\rm out, AGB}=10 {\rm\thinspace km~s^{-1}}$, following the typical observed outflowing velocities of AGB driven winds \citep[e.g.][]{Nyman1992}.

        In this model, various mass-loss events such as winds from young stars, SNe, and AGB stars provide recycled gas, which can fuel late star formation and also AGN activity by feeding the central supermassive black holes \citep[e.g.][]{Ciotti2010}. With an assumed \citet{Kroupa2001} IMF, over 30\% of the total mass in stellar particles returns to the ISM within $\sim 13$ Gyr of evolution.

    \subsection{Black Hole Physics and AGN Feedback}
        
        In the simulations, the black holes are treated as collisionless sink particles and seeded with a mass of $10^5 \thinspace h^{\rm -1} { \rm M_{\odot}}$ in newly formed dark matter halos with mass above $10^{11} \thinspace h^{\rm -1} {\rm M_{\odot}}$. The black hole can grow via mergers with other black holes and direct gas accretion. Two black holes merge when they are within their local SPH smoothing lengths, and their relative velocities are less than the local sound speed. The rate of the gas infall onto the black hole is estimated with a Bondi-Hoyle-Lyttleton parameterization \citep{Hoyle1939,Bondi1944,Bondi1952}: $\dot{M}_{\rm{inf}}= (4 \pi  G^{2} M_{\rm BH}^{2} \rho ) /((c_{\rm s}^2+ v^{2})^{3/2})$, where $\rho$, $c_{\rm s}$, and $v$ are respectively the gas density, the sound speed and the velocity of the gas relative to the black hole. The simulations also include the soft Bondi criterion introduced in \cite{Choi2012} to prevent the unphysical infall of gas outside the Bondi radius. This model statistically limits the accretion to the gas within the Bondi radius.
        
        Black hole modelling also includes the mechanical and radiative AGN feedback model described in \citet{Choi2012,Choi2015}. In this model, the BH imparts mass and momentum to the surrounding gas in a manner that mimics observed AGN-driven outflows \citep[e.g.][]{Arav2020}. As AGN winds carry a mass, only a fraction of inflowing gas mass will accrete onto the black hole as $\dot{M}_{\rm outf} = \dot{M}_{\rm inf} - \dot{M}_{\rm acc}$ where $\dot{M}_{\rm outf}$, $\dot{M}_{\rm inf}$ and $\dot{M}_{\rm acc}$ are the outflowing/inflowing mass rate and the mass accretion rate to the black hole respectively. The model assumes a constant AGN-driven wind velocity of $v_{\rm outf,AGN} = 10,000 {\rm\thinspace km~s^{-1}}$, and the wind energy is calculated as $\dot{E}_w  \equiv \epsilon_{\rm w} \dot{M}_{\rm acc} c^2= 0.5 \dot{M}_{\rm outf} v_{\rm outf,AGN}^2$, where  the feedback efficiency parameter is set to $\epsilon_{\rm w} = 0.005$ to reproduce the present day $M_{\rm BH}-\sigma$ relation \citep{Choi2017}. Then the mass flux and momentum flux carried by the AGN-driven winds are calculated as $\dot{M}_{\rm outf} = 2 \dot{M}_{\rm acc} \epsilon_{\rm w} c^2 / v_{\rm outf,AGN}^2 $ and $\dot{p} = \dot{M}_{\rm outf} v_{\rm outf,AGN} = 2 \epsilon_{\rm w} \dot{M}_{\rm acc} c^2 / v_{\rm outf,AGN}$. In short, when the inflow of gas feeds the central black hole, it launches fast winds carrying a momentum flux of $\dot{p} = 30 L_{\rm BH} / c$. The direction of the wind is set to be parallel or anti-parallel to the angular momentum vector of each gas particle accreted by the black hole to mimic a wind perpendicular to the disk plane.

        The heating and associated radiation pressure effect from moderately hard X-ray radiation ($\sim 50$~keV) from the accreting black hole are also included. Heating and radiation pressure follow the AGN spectrum and Compton/photoionization heating prescription from \citet{Sazonov2004,Sazonov2005}, including the associated radiation pressure directed away from the black hole. In these simulations, super-Eddington gas accretion occasionally occurs as the accretion rate onto the black holes is not artificially capped at the Eddington rate. Instead, the simulations include an Eddington force acting on electrons in gas particles, directed radially away from the black hole so that the corresponding feedback effects can reduce the gas inflow.

\section{Methods} \label{Methods}

    We now walk through the various methods used throughout this paper. Throughout our analysis, we refer to ``intrinsic'' properties, which are extracted directly from the simulations, and ``synthetic'' properties, which are calculated from the mock-observed images.
    
    We focus on galaxies within the redshift range $0.5 < z < 3$ where CANDELS can well-resolve massive galaxies \citep{Barro2019}, mergers are common \citep[e.g,][]{OLeary2021}, and AGN are common within CANDELS \citep[e.g,][]{Kocevski2012}. For each $29$ halo, we select the $43$ snapshots spanning our redshift range of interest, yielding $1247$ snapshots. We further select snapshots where the halo-finder can accurately track the main halo, paring the total down to $1156$ simulation snapshots. Using a friends-of-friends algorithm on gas + star particles, we identify stellar companions surrounding the main galaxy. Next, we measure the intrinsic properties of mergers and AGN activity within the simulation snapshots. We then use those snapshots to generate mock observations and measure morphological parameters from each mock observation. Finally, we train a machine learning classifier on the intrinsic merger information to measure merger information from the mock observations. This classifier allows us to reasonably compare mergers derived from the mock observations and those intrinsic to the simulation.
    
    Throughout our analysis, we compare to CANDELS data from \citet{Kocevski2012,Rosario2015} out to $z < 3$, which include galaxy morphologies measured in the F160W band from the EGS, UDS, GOODS-S, and GOODS-N fields. X-ray follow-up in all but the UDS CANDELS fields comes from publicly available Chandra point-source catalogues. They use the 4 Ms \citep{Xue2011} and 2 Ms \citep{Xue2016} point-source catalogues for GOODS South and North, and the 800 ks catalogue \citep{Nandra2015} for EGS. For UDS, they use a 600 ks source catalogue from the X-UDS survey \citep{Kocevski2018}. We also compile stellar masses from catalogues for GOODS South and UDS \citep{Santini2015}, EGS \citep{Stefanon2017}, and GOODS North \citep{Barro2019}. When directly comparing to CANDELS, we mass-match CANDELS galaxies to match our simulated stellar masses.
    
    \subsection{Identifying Intrinsic Mergers} \label{merger-identification}
        
        Observations of AGN within merging systems find the most luminous AGN in the latest stages of major mergers \citep{Ellison2013, Koss2018}. If galaxy mergers fuel AGN activity, we expect a time delay between companion coalescence and the start of AGN activity. The observability timescales of merger features have been found to be between $0.2 - 0.4$ Gyr \citep{Lotz2010,Ji2014}, though \citet{Mantha2018} find that observability timescales evolve with redshift. 

        We identify mergers in the simulation using a close pairs method similar to \citet{Ferreira2020}, in which stellar companions above a stellar mass ratio, $\M{1}/\M{2}$, fall within some 3D physical distance, $R$, of the primary galaxy's centre. Furthermore, we impose the additional criteria that companions have low 3D velocities relative to the main galaxy $v_{\rm rel} < 500$ km s$^{-1}$ \citep{Mantha2018,Ventou2019}. Starting at $z=0.5$ in each zoom, we trace the primary halo's merger tree backward in cosmic time and search for snapshots where a companion meets our merging criteria. We label snapshots where at least one valid close pair is found as the central snapshot of the merger event, $t_{\rm pair}$. Then, all earlier+later snapshots within a chosen timescale are also marked as part of the merger event.
        
        In this work, we calculate mass ratios by taking the stellar mass of companions at the time close pairs are identified. However, \citet{Rodriguez-Gomez2015} find that calculating mass ratios based on companion stellar mass prior to infall helps mitigate the various effects of merging on stellar mass (e.g., through stripping or induced star formation). Although our method better matches how close pairs are identified by observers and side-steps the complexities of tracing the stellar progenitor history, our method is also impacted by the companion-finder's ability to correctly measure companion stellar masses at small separations. We expect the number of detectable minor mergers to be significantly reduced relative to some other merger identification methods. As such, we limit our explored mass ratios to only major mergers, $\M{1}/\M{2} \geq 0.25$.  According to our definition of major merger, 21 simulated galaxies out of 29 have at least one major merger event in their lifetime.
        
        In determining a pair separation, we analyzed common separations from the literature \citep{Lotz2010, Bignone2017}. However, maximum separations much smaller than $R < 50$ kpc capture very few mergers, likely because the time between simulation snapshots is too long (between $130-160$ Myr).
        
        In choosing a merging timescale, we first set a fiducial timescale of $\pm 300$ Myr from the central snapshot as in \citet{Ferreira2020}. We then explored various other timescales, including shorter/longer ranges, asymmetric ranges, and redshift-dependent ranges \citep{Mantha2018}. However, 1) the fiducial timescale produces merger fractions similar to observations (see Section \ref{true-merger-statistics}), and 2) other timescales all performed equally well or worse in our automated classifier. Therefore, we omit results using these other time ranges for clarity and only show results for a merging timescale of $\pm 300$ Myr.
    
        The parameters we explore in this work are summarized as follows:
        \begin{itemize}
            \item $\M{1}/\M{2} \geq 0.25$,
            \item $R < 50$ kpc,
            \item $[t_{\rm pair} - 300, t_{\rm pair} + 300]$ Myr.
        \end{itemize}

    \subsection{Synthetic Image Generation}
    
        \begin{figure*}
            \includegraphics[width=\linewidth]{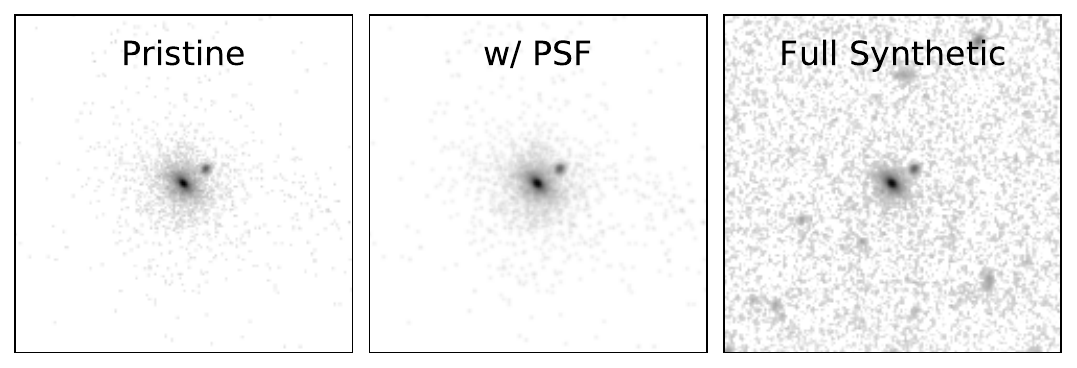}
            \caption{Each step of the synthetic image generation pipeline for a randomly selected synthetic image. Each panel is cropped to a physical scale of $30$ kpc across for clarity. The left panel shows the pristine F160W image output for halo \textit{m0215} at $z=1.5$ from the RT code \textsc{Powderday}. The centre panel shows the image after being properly redshift dimmed, being PSF-convolved, and having source shot-noise added. The right panel shows the image after injecting into a randomly selected CANDELS field. The right panel is a cropped example of our mock observations, from which we extract morphological parameters.}
            \label{pipeline-comparison}
        \end{figure*}
    
        \subsubsection{Radiative Transfer}
        
            We generate noiseless, high-resolution images from our galaxy simulation data using the open-source radiative transfer (RT) code {\sc Powderday} \citep{Narayanan2021}, which combines dust RT from {\sc Hyperion} \citep{Robitaille2011} with stellar population synthesis models from \textsc{FSPS} \citep{Conroy2009,Conroy2010}, and uses {\sc yt} \citep{Turk2011} as the underlying simulation loader. Each star particle is assigned a spectral energy distribution (SED) according to its mass, age, and stellar metallicity, with an underlying \citet{Kroupa2001} initial mass function. Similarly, each black hole particle is assigned an empirical \citet{Hopkins2007} SED according to its bolometric luminosity. These SEDs are then projected through the interstellar medium to generate broadband $(912$\AA$ - 1$ mm$)$ galaxy SEDs. We use {\sc Powderday} to form a dust grid with a constant dust-to-metal ratio of $0.4$ and a \citet{Draine2003} Milky Way dust model, then generate galaxy SEDs as well as filter-convolved photometric images. 
            Each of our $1156$ galaxy snapshots is ``observed" at $3$ orientations randomly selected with respect to the galaxy and run through {\sc Powderday}. This process generates $3468$ total images and SEDs, properly redshifted to the cosmic time of the snapshot. We then convolve the output SEDs with the HST-WFC3 F160W filter at an image resolution of $512 \times 512$ pixels. Each image has a physical scale of $200$ kpc, yielding pixel sizes of $390$ pc. Every star and gas particle within the viewing window, including those from companion galaxies, is included in the RT. The angular size does not strongly change between $0.5 < z < 3$, so we keep a constant image size at all redshifts. We apply a cosmological surface brightness dimming by a factor of $(1+z)^{-5}$ to each image, in the original units of erg s$^{-1}$ cm$^{-2}$ nm$^{-1}$ Sr$^{-1}$.

        \subsubsection{Simulation of mock HST images}
        
            To convert our ideal sky images into HST-like synthetic images, we follow a process similar to \citet{Snyder2015a,Snyder2019} by convolving with a point-spread function (PSF) and then introducing noise. The $390$ pc pixel size of the ideal sky images is approximately $0.05$ arcsec at $0.5 < z < 3$, similar to the WFC3-IR mosaic pixel sizes ($0.06$ arcsec) of the CANDELS drizzled images, and we choose PSF parameters to approximate the quality of those data products \citep{Koekemoer2011,Grogin2011}.

            We first convolve each pristine image with a Gaussian PSF with full-width at half-maximum of $0.145$ arcsec, corresponding to the $0.145$ arcsec core PSF of HST-WFC3 in the F160W filter. Then, we add source shot noise to each pixel of the PSF-convolved images.
            
            Next, we inject each of our PSF-convolved + noise-added images into a randomly selected CANDELS image. We use a modified version of the \textsc{RealSim} package \citep{Bottrell2019} to generate what they label ``fully realistic images". They find that convolving synthetic images into real fields dramatically improves the performance of automated merger classifiers. We follow a similar prescription to \citet{Bottrell2019}:
            \begin{enumerate}
                \item Randomly select a CANDELS field, choosing from the EGS, UDS, COSMOS, and GOODS-S fields. We omit GOODS-N because of its different F160W photometric resolution.
                \item Randomly select a galaxy from the corresponding EGS \citep{Stefanon2017}, UDS \citep{Galametz2013}, COSMOS \citep{Nayyeri2017}, or GOODS-S \citep{Guo2013} full multiband catalogues.
                \item Generate a deblended segmentation map for the F160W image, within a $1024 \times 1024$ cutout centreed on the selected galaxy.
                \item Randomly select an injection site within the cutout, with the restriction that the centre of the injected image does not land on another object in the segmentation map.
                \item Inject the mock image at the selected site, and extract the final $512 \times 512$ synthetic image cutout.
            \end{enumerate}
            Figure \ref{pipeline-comparison} illustrates each step of the synthetic image generation pipeline for a randomly selected image.
            
    \subsection{Observational Merger Classifiers}
    
        Using mock observations, we measure morphological statistics commonly used to identify mergers. We calculate the Gini $(G)$ and $\M{20}$ statistics \citep{Abraham2003,Lotz2004}, as well as the asymmetry $(A)$ and concentration $(C)$ statistics \citep{Conselice2003} using the {\sc statmorph} package \citep{Rodriguez-Gomez2019}. We train an automated machine learning classifier with these morphological parameters as features to reproduce the intrinsic merger statistics.
        
        Before measuring morphological statistics, we background subtract each of our images and generate deblended segmentation maps using {\sc photutils}. We first estimate the background by iteratively measuring the 3$\sigma$ clipped median pixel flux until convergence. Then, we measure the morphologies of sources with at least $16$ connected pixels $1.5\sigma$ above the background.
        
        
        \subsubsection{Morphological Measurements}
        
            \begin{figure*}
                \includegraphics[width=\linewidth]{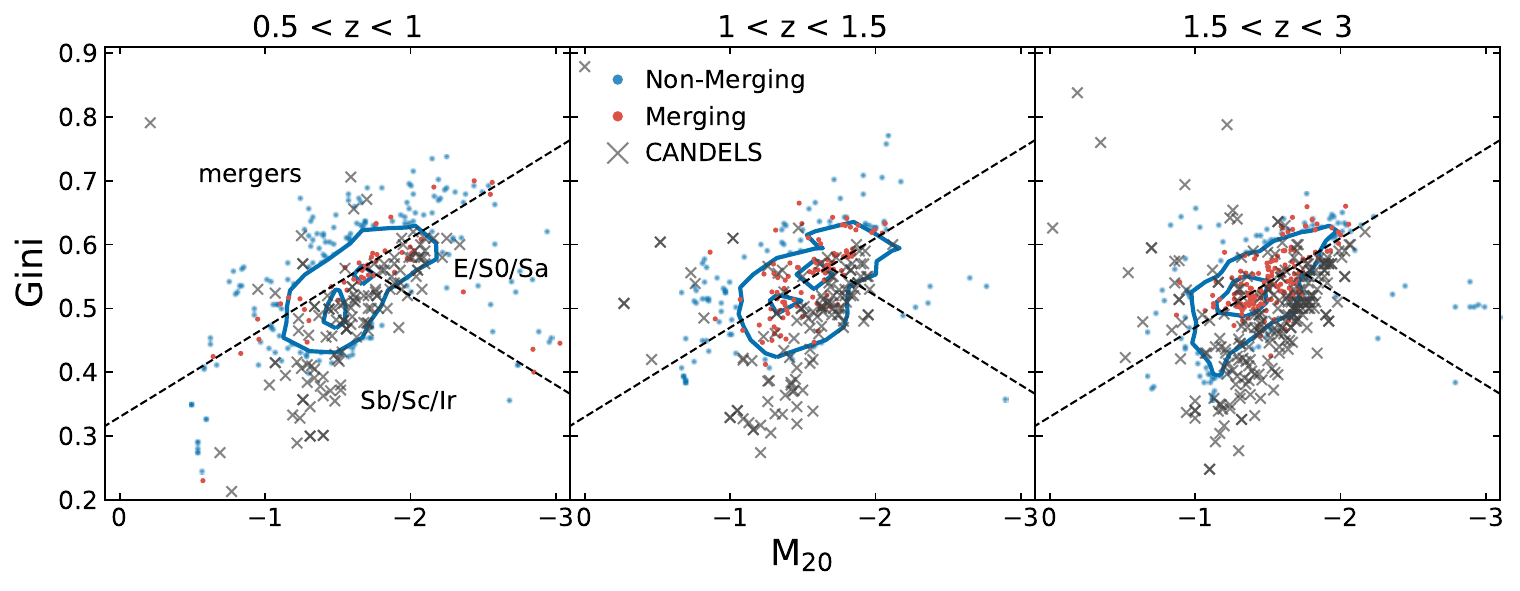}
                \caption{Gini versus $\M{20}$ for synthetic images and mass-matched CANDELS data, organized into redshift bins. Dashed lines show \citet{Lotz2008} morphological divisions between merging versus non-merging, and E/S0/Sa versus Sb/Sc/Irregular galaxy morphologies. Synthetic image morphologies for both non-mergers (blue contours, points) and major mergers (orange points) tend to lie along the typical dividing line $S(G,\M{20}) = 0$. Synthetic image morphologies are slightly higher in the Gini index than CANDELS morphologies (black $\times$).}
                \label{G-M20}
            \end{figure*}
        
            \begin{figure*}
                \includegraphics[width=\linewidth]{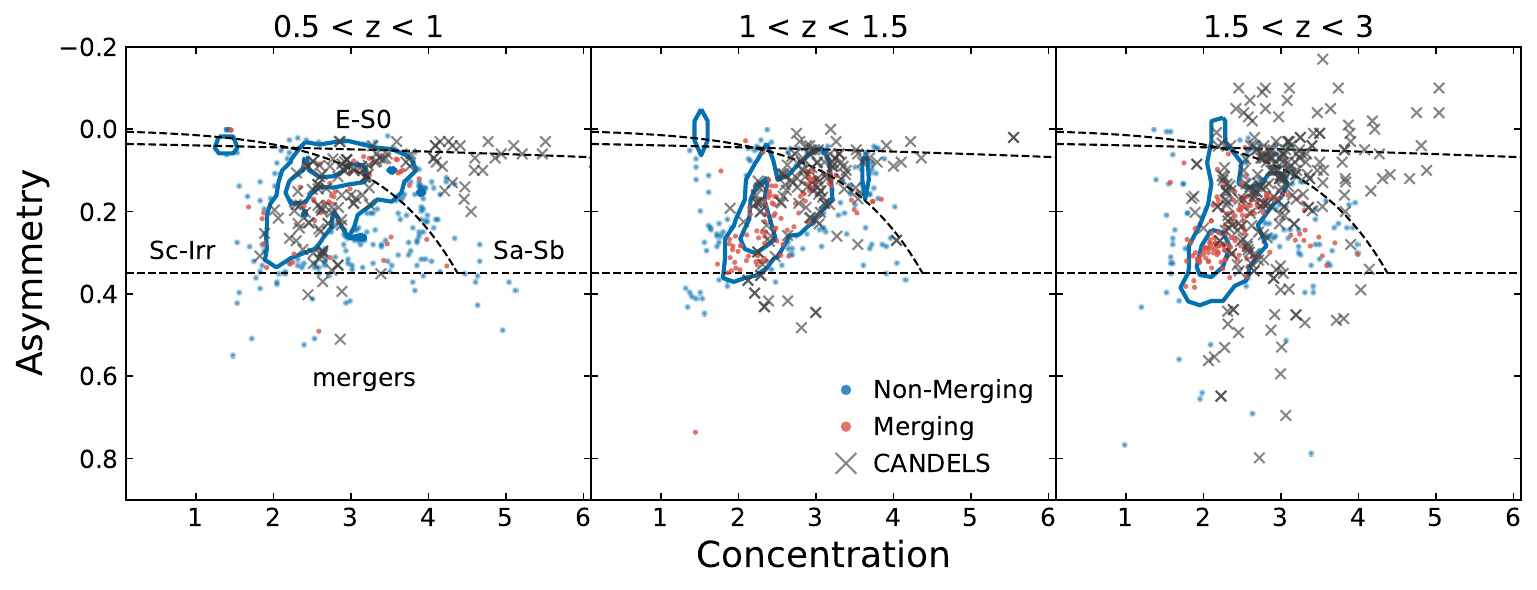}
                \caption{Asymmetry versus concentration for synthetic images and mass-matched CANDELS data, organized into redshift bins. Dashed lines indicate typical boundaries for mergers \citep[$A > 0.35$,][]{Conselice2003}, as well as the typical boundaries between E/S0, Sa/Sb, and Sc/Irregular morphologies \citep{Bershady2000}. Synthetic image morphologies for non-mergers (blue contours, points) and mergers (orange points) lie mostly in the late-type region (Sa/Sb/Sc/Irr), with a small fraction lying in early-type regions (E/S0) and merging galaxy morphologies. CANDELS (black diamonds) tends to exhibit more early-type morphologies, but occupy approximately similar regions of the space.}
                \label{ASYM-CONC}
            \end{figure*}

            We measure non-parametric morphological statistics designed to detect the presence of disturbance or multiple nuclei. We use {\sc statmorph} to calculate $G$, $\M{20}$, $A$, and $(C)$, among other statistics. Images that fail any morphological fit are automatically flagged by {\sc statmorph}, which we discard for purity. Approximately $4$\% of images are flagged and discarded, leaving us with $3327$ synthetic images with valid morphological fits.
        
            Gini's coefficient, $G$, measures the inequality among pixel flux values and is normalized such that it ranges from 0 (all pixels contain equal flux) to 1 (one pixel contains all flux). $G$ is location-independent and hence is able to detect multiple bright nuclei but is sensitive to which pixels are selected for measurement. Building off the original construction in \citet{Glasser1962}, \citet{Lotz2004} find that $G$ can be robustly calculated for galaxies above signal-to-noise $S/N > 3$:
    
            \begin{equation}
                G = \frac{1}{\left|\bar{X}\right|n\left(n-1\right)} \sum^n_{i=1}\left(2i - n - 1\right)\left|X_i\right|,
            \end{equation}
            where all $n$ pixels are rank-ordered by absolute flux values $\left|X_i\right|$, and $\left|\bar{X}\right|=\sum_i \left|X_i\right|/n$ is the mean absolute flux value.
    
            The $\M{20}$ statistic calculates the second-order moment of the brightest quintile of pixel flux values, relative to the total second-order moment, within the segmentation map \citep{Lotz2004}:
    
            \begin{equation}
                \M{20} = \log \frac{\sum_i M_i}{M_{\rm tot}}{\rm, \quad for } \sum_i I_i < 0.2 I_{\rm tot},
            \end{equation}
            where
            \begin{equation}
                M_{\rm tot} = \sum^n_i I_i \left[\left(x_i - x_c\right)^2 + \left(y_i - y_c\right)^2\right],
            \end{equation}
            and $x_c$, $y_c$ are the galaxy centre coordinates defined to minimize $M_{\rm tot}$.
    
            The $C$ statistic is typically defined:
            \begin{equation}
                C = 5\log_{10} \left(\frac{r_{80}}{r_{20}}\right),
            \end{equation}
            where $r_{20}$ and $r_{80}$ are the circular aperture radii containing $20$\% and $80$\% of the galaxy flux within $1.5\times$ the Petrosian radius. The aperture centre is determined by the point that minimizes the $A$ statistic:
            \begin{equation}
                A = \frac{\sum_{i,j}\left|I_{ij} - I_{ij}^{180}\right|}{\sum_{ij}\left|I_{ij}\right|} - A_{\rm bgr},
            \end{equation}
            where $I_{ij}$ and $I_{ij}^{180}$ are the pixel fluxes for the image and image rotated by $180^{\circ}$, respectively. $A_{\rm bgr}$ is the average background asymmetry. The pixel flux values within $1.5 r_{\rm petro}$ are included in the computation.
                
            Following \citet{Snyder2019}, we rotate the $G - \M{20}$ space to define two simplifying diagnostics relative to the locus of points found in \citet{Lotz2008}, the bulge statistic $F(G, \M{20})$ and the merger statistic $S(G, \M{20})$:
    
            \begin{equation}
                F(G,\M{20}) = 
                    \begin{cases}
                        |F| & G \geq -0.14 \M{20} + 0.778\\
                        -|F|, & G < -0.14 \M{20} + 0.778,
                    \end{cases}
            \end{equation}
            where $|F| = |-0.693 M_{20} + 4.95G - 3.85|$ \citep{Snyder2015a}. $F$, which extends parallel to the locus of points, is less sensitive than $\M{20}$ to the effects of dust and mergers, which tend to move galaxies perpendicular to the locus of points. $S$ instead extends perpendicular to the locus of points:
    
            \begin{equation}        
                S\left(G,\M{20}\right) = 
                    \begin{cases}
                        |S| & G \geq -0.14 \M{20} + 0.33\\
                        -|S|, & G < -0.14 \M{20} + 0.33,
                    \end{cases}
            \end{equation}
            where $|S| = |0.139 M_{20} + 0.990G - 0.327|$ \citep{Snyder2015}. Positive values tend to identify multiple cores (high $\M{20}$) and starbursts (high $G$).

            Figure \ref{G-M20} shows the distribution of the data from the synthetic images in $G-\M{20}$ space in bins of redshift. Our morphologies tend to lie on or below the \citet{Lotz2008} dividing line between mergers and non-mergers and exhibit slightly higher values of $G$ relative to CANDELS galaxies. The merger dividing line corresponds with the line of $S(G, \M{20}) = 0$. At high redshift, disk-dominated galaxies are more prevalent than bulge-dominated galaxies ($42\%$ versus $23\%$, respectively). At low redshift, disk-dominated galaxies become rarer than buldge-dominated ($8\%$ versus $46\%$, respectively). The fraction of galaxies merging according to simple cuts on $S(G, \M{20})$ varies between $35 - 47\%$ . In CANDELS, disk-dominated galaxies are always more prevalent than bulge-dominated ($\sim60\%$ versus $\sim20\%$), highlighting that our cosmological zoom simulations are not cosmologically representative.

            Figure \ref{ASYM-CONC} compares our morphologies to mass-matched CANDELS galaxies in asymmetry-concentration space. We include \citet{Bershady2000} divisions between bulge- and disk-dominated galaxies and merging systems. Between $90-100\%$ of synthetic morphologies lie within late-type regions, quite different from the $65-85\%$ of CANDELS galaxies. Mergers identified through a single cut in asymmetry make up a small fraction of the sample between $0-3\%$, different from CANDELS' $2-12\%$. Despite these differences, synthetic image morphologies occupy similar regions of asymmetry-concentration space as CANDELS.
            
            Notably, in both $G-\M{20}$ space and $A-C$ space, true mergers are not strongly morphologically distinct from non-mergers. These distributions illustrate why flat cuts on a single morphological diagnostic will not strongly distinguish mergers, and why a machine learning classifier may be necessary to identify mergers.

            \subsubsection{A new automated method to select mergers} \label{automated-classifier}
                
                \begin{figure} 
                    \includegraphics[width=\linewidth]{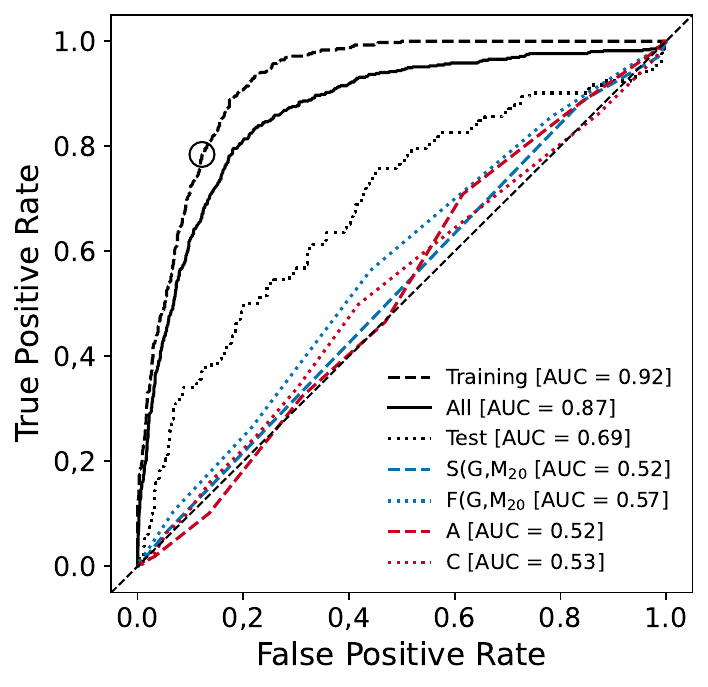}
                    \caption{ROC curve for the RF classifier trained on the synthetic image training set (black dashed), test set (black dotted), and applied to the total data set (black solid). The training and test galaxies are randomly chosen at a $2/3$ and $1/3$ split, respectively. The F1-scored probability threshold calculated on the validation set is marked by a black circle. We also show ROC curves recovered from flat cuts in $S(G, \M{20})$ (blue dashed), $F(G, \M{20})$ (blue dotted), asymmetry (red dashed), and concentration (red dotted). We include area-under-curve (AUC) values for each curve. The RF classifier shows  better ROC performance than cuts on a single morphological diagnostic.}
                    \label{roc-curve}
                \end{figure}
                
                \begin{figure} 
                    \includegraphics[width=\linewidth]{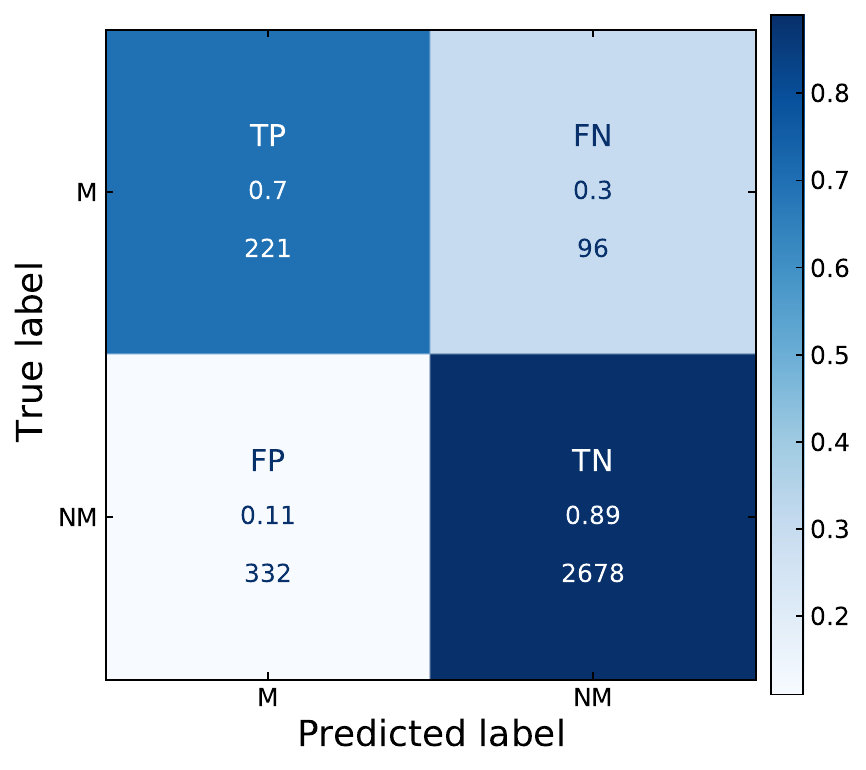}
                    \caption{Confusion matrix for the trained RF classifier. The vertical axis marks true mergers and non-mergers determined directly from the simulation, while the horizontal axis marks the tuned classifier's prediction of mergers and non-mergers. The confusion classes are true positive (top left, intrinsic mergers selected), false negative (top right, mergers rejected), false positive (bottom left, non-mergers selected), and true negative (bottom right, non-mergers rejected). Each panel indicates the confusion class, the fraction of true mergers/non-mergers within the class, and the number of images within the class. Panels are coloured according to the stated fraction of true mergers/non-mergers that reside within the classification bin. The classifier is able to identify $70\%$ of true mergers, at the cost of misidentifying $11\%$ of non-mergers as mergers. Since non-mergers are much more common than mergers, the precision of the classifier is low at $40\%$.}
                    \label{confusion-matrix}
                \end{figure}
                
                \begin{figure*}
                    \includegraphics[width=\linewidth]{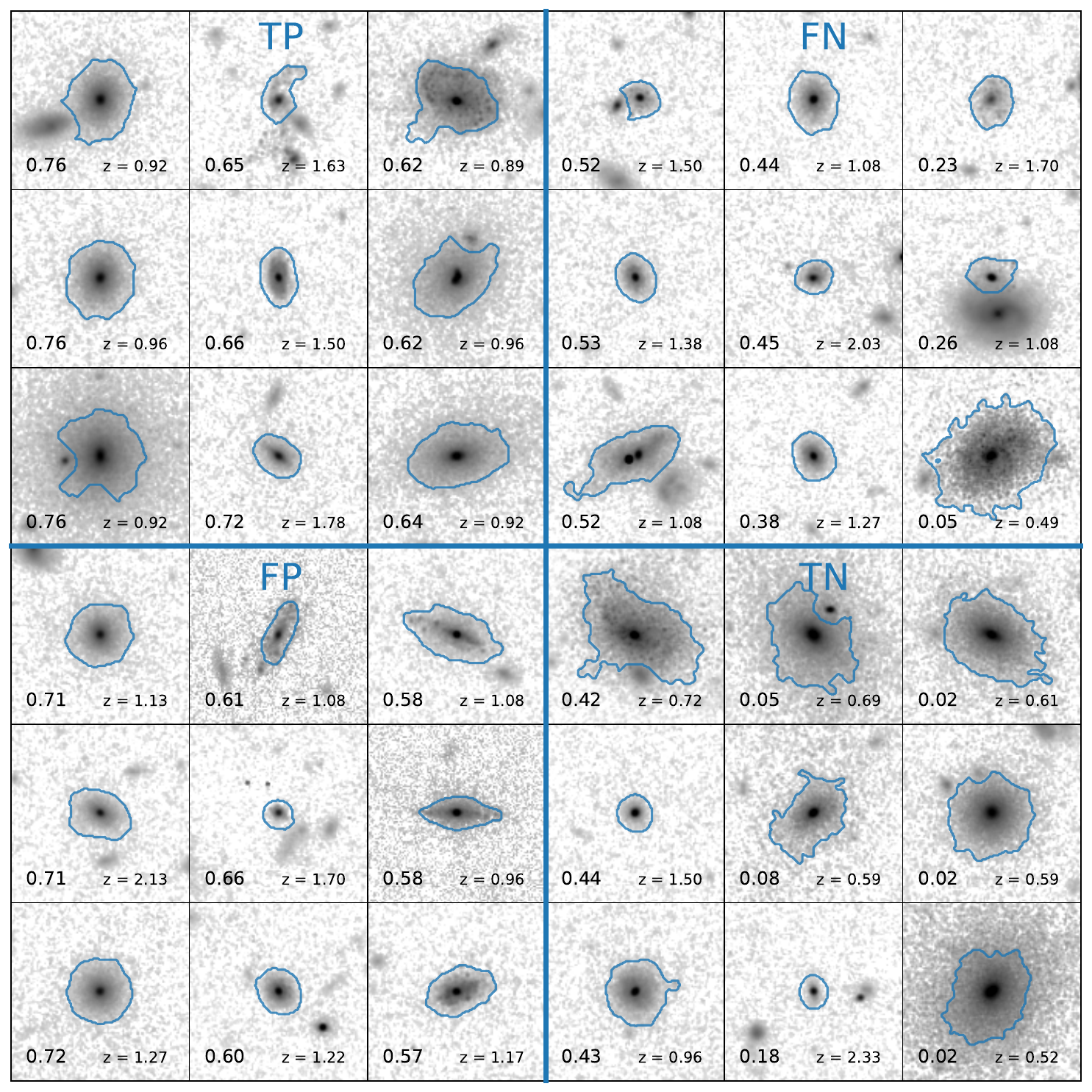}
                    \caption{Randomly selected F160W mock observations grouped by confusion class. Each panel includes the classification probability output by the RF classifier at the maximal $f1$ point, as well as the redshift of the target galaxy. We include the deblended segmentation maps used to calculate morphologies (blue contours) to aid the eye in determining where deblending may have contributed to false classifications.}
                    \label{synthetic-images}
                \end{figure*}
                
                Simple cuts on $S\left(G, \M{20}\right)$ have been found to separate many $z < 0.25$ systems with bright nuclei and bright tidal tails, which together can indicate recent or ongoing merger activity \citep{Lotz2004}. Similarly, simple cuts on asymmetry have been found to identify highly disturbed systems out to $z=3$ \citep{Conselice2003}. However, we find that simple cuts on synthetic image morphologies do not accurately identify intrinsic mergers. Instead, we pursue an automated approach to merger classification, which gives more accurate merger classifications.
            
                There are a number of approaches to automated merger classification based on either the multidimensional space of morphological diagnostics or the image pixel values themselves. \citet{Nevin2019} perform linear discriminant analysis on a seven-dimensional morphological diagnostic parameter space to generate a classifier that robustly classifies major and minor mergers separately, but not together. Both \citet{Goulding2018} and \citet{Snyder2019} use a random forest ensemble learning method to classify mergers based on a number of parametric and non-parametric diagnostics, including those described in this work. Such approaches allow for direct comparisons with diagnostic-based observations at the potential cost of missing merger-relevant features not present within the diagnostics. Other studies such as \citet{Huertas-Company2018, Huertas-Company2019,Bottrell2019,Ferreira2020,Ciprijanovic2020,Bickley2021} use deep-learning methods to robustly identify morphological features from image pixel values themselves, at the cost of large training set size requirements and greater difficulty in direct comparisons with other work.
        
                We choose a random forest (RF) ensemble learning method for classifying mergers from our synthetic images. RFs are a non-parametric method that generates decision trees to subdivide the multidimensional parameter space. At each step of a given tree, a feature subdivision is chosen by identifying the subdivision with the largest reduction in entropy. RFs benefit from their ability to be trained on high-level summary statistics rather than full astronomical images. However, as with many other methods, they require careful tuning of the many hyperparameters. 
                
                When trained on unbalanced training sets, random forests may also present biases towards the majority class. Since intrinsic mergers are rare relative to non-mergers, we perform a random undersampling of the majority class (non-mergers) to bring the two classes into balance. Results change little when we instead apply random oversampling of the minority (mergers) class. However, these sampling techniques are not perfect, and any ML classifier will suffer in classification precision when trained on strongly imbalanced data.
                
                We use the \textsc{Scikit-learn} package \citep{Pedregosa2011} to train and test a RF classifier on the non-parametric morphological statistics extracted from our synthetic images ($G$, $\M{20}$, $A$, and $C$) as well as redshift. We label each image with a True or False binary classification based on the given intrinsic merger definition. For simplicity, we show results for a single intrinsic merger definition that performed the best (major pairs with separations of at most $50$ kpc). Other intrinsic merger definitions exhibit worse classification performance. We train the RF on the images from a random subset of $2/3$ of the galaxies, then test classifications, and report results on images from the remaining, untouched $1/3$ of galaxies. Splitting the training/test sets in this way ensures that correlations between subsequent snapshots of the same galaxy do not leak from the training set into the test set. We use the \textsc{Imbalanced-learn} package \citep{Lemaitre2017} to perform random oversampling of the minority class during the bagging step within the cross-validation pipeline. Oversampling is performed on the training folds, and then the classifier is scored on the non-oversampled validation fold. We use $10$-fold cross-validation with $500$ decision trees up to a maximum depth of $50$ layers, up to $4$ features used per split, and up to $6$ minimum samples per leaf. The tuned classifier reaches a maximum tree depth of $32$ layers, $2$ parameters per split, and a minimum of $4$ samples required to be at a leaf node. The RF outputs the classification probability for each image, which we transform into a classification prediction.
        
                In order to map classification probabilities to binary classifications, we must first define common classification measures:
                \begin{itemize}
                    \item True Positive (TP): True mergers selected by classification.
                    \item False Positive (FP): Non-mergers selected by classification.
                    \item True Negative (TN): Non-mergers rejected by classification.
                    \item False Negative (FN): True mergers rejected by classification.
                \end{itemize}
        
                Using these measures we also define the True Positive Rate (or recall):
                \begin{equation}
                    TPR = \frac{TP}{TP + FN},
                \end{equation}
                the False Positive Rate:
                \begin{equation}
                    FPR = \frac{FP}{FP + TN},
                \end{equation}
                and the Positive Predictive Value (or precision):
                \begin{equation}
                    PPV = \frac{TP}{TP + FP}.
                \end{equation}
        
                A classifier that perfectly includes mergers and excludes non-mergers would hence have a $TPR = 1$, $FPR = 0$, and $PPV = 1$. We select a classification probability threshold using the $F_1$ statistic, which identifies the probability threshold at which:
                
                \begin{equation}
                    F_1 = \frac{TP}{TP + \frac{1}{2}(FP + FN)},
                \end{equation}
is maximized. The $F_1$ statistic attempts to strike a balance between precision and recall. Scoring based on precision and recall is especially important in unbalanced data sets like ours, which has fewer intrinsic mergers than non-mergers.
                
                Figure \ref{roc-curve} shows the Receiver Operating Characteristics (ROC) curve for the RF-trained classifier, which measures the diagnostic ability of the classifier as the decision boundary changes. We show the ROC curve and AUC values for the training set as well as the total data set, marking the maximal $F_1$ probability threshold on the validation set with a circle. We also include the ROC curves for classification based solely on sliding $S(G, \M{20})$, $F(G, \M{20})$, asymmetry, and concentration thresholds. The AUC value for the all data is 0.87 and the AUC value for the test data is 0.69. Considering the ROC curve on the test dataset, the RF classifier has reasonable discrimination performance. Most importantly, the RF classifier outperforms the simple classifiers, whereas the simple classifiers perform slightly better than a random selection.
                
                Figure \ref{confusion-matrix} shows the confusion matrix for the tuned classifier applied to the full data set. In each classification bin, we include the number of images as well as the fraction of true mergers/non-mergers (i.e., the fraction of images in each row) that fall within each bin. Notably, the classifier falsely identifies many non-mergers as mergers, lowering the precision.
        
                Classifying mergers in this way has a few drawbacks. Foremost, the precision of the RF classifier is low, where only $40$\% of positive classifications are true positives. As seen in the confusion matrix, although $70\%$ of true mergers are correctly identified, $11\%$ of non-mergers are falsely classified as mergers. Our synthetic images have been specifically generated to emulate CANDELS-like observations, and our observational merger diagnostics likely perform differently at different image resolutions or depths. Hence our classifier likely does not generalize well outside of the specific mock observation environment.
                
                Figure \ref{synthetic-images} shows randomly selected synthetic images after classification, along with the deblended segmentation maps used for morphological analysis. Each image is cropped to a physical scale of $50$ kpc and indicates the target object redshift along with the classification probability. We include segmentation maps to aid the eye in identifying where deblending may have contributed to false classification.

\section{Results} \label{Results}

    We now search for any connection between mergers and triggering of AGN activity by 1) analyzing properties of intrinsic mergers, 2) analyzing properties of mergers classified from the synthetic images, and 3) comparing intrinsic, classified, and observational merger statistics. When we examine intrinsic mergers in our following analysis, we only focus on widely separated ($< 50$ kpc) pairs with a major companion ($>1:4$) and low relative velocities ($v_{\rm rel} < 500$ km s$^{-1}$). Snapshots within $\pm 300$ Myr of $t_{\rm pair}$ are also marked as mergers.
    
    A vital aspect of our analysis is accurately estimating the luminosities of our SMBHs. We assume a piecewise expression for bolometric luminosity that depends on the Eddington ratio, $f_{\rm Edd}$, such that radiatively inefficient BHs have reduced luminosities:
    
    \[
    L_{\rm bol} = 
        \begin{cases} 
          \frac{\epsilon_r}{1-\epsilon_r} \dot{M}_{\rm BH} c^2 & f_{\rm Edd} > 0.1 \\
          10 f_{\rm Edd} \epsilon_r \dot{M}_{\rm BH} c^2 & f_{\rm Edd} < 0.1, \\
        \end{cases}
    \]
    where $\dot{M}_{\rm BH}$ is the accretion rate of the SMBH, and we choose $\epsilon_r = 0.1$ \citep{Churazov2005,Hirschmann2014,Habouzit2019}. We also adjust our AGN luminosities to account for AGN variability on timescales below the simulation resolution limit. \citet{Hickox2014} find that models, where AGN spend more time at low luminosities fractionally, are able to reproduce observed relationships between AGN luminosity and star formation rate. We adopt their fiducial model, which follows a Schechter function of the form:
    
    \begin{align}
        \frac{dt}{d\log L} = t_0 \left(\frac{L}{L_{\rm cut}}\right)^{-\alpha} \exp\left(-L/L_{\rm cut}\right),
    \end{align}
    where $L$ is a relative luminosity, $\alpha = 0.2$, and $L_{\rm cut} = 100\left<L_{\rm bol}\right>$. For each of our galaxies, we sample the distribution and extract a new $L_{\rm bol}$. Including short-timescale variability adds additional variance to our AGN luminosities, with a longer tail towards dim SMBHs.
    
    Throughout our analysis, we frequently compare the properties of galaxies to the properties of appropriately chosen control galaxies. There are many ways to choose control galaxies, including selecting based on redshift, galaxy physical properties, or environment. In this work, we select according to stellar mass, redshift, and an additional criterion based on the quantity we wish to explore. For example, in studying the prevalence of mergers in AGN, our sample of AGN with bolometric luminosities $L_{\rm bol} > 10^{43}$  \ergs\ is paired with inactive control galaxies with $L_{\rm bol} < 10^{43}$ \ergs. In general, to generate a control sample, we first choose suitable control criteria, then select control galaxies within a factor of $2$ of the target galaxy's stellar mass from snapshots within $\pm 300$ Myr of the target galaxy's snapshot. Each target galaxy is paired with $2$ control galaxies. We find that strengthening or relaxing any of these criteria does not strongly affect results, though it can influence the scale of our uncertainties in cases where few suitable control galaxies are found.
    
    \subsection{Properties of Intrinsic Mergers}
    
        We begin by exploring the properties of intrinsic mergers obtained from the simulation. We examine the merger fractions predicted by the simulation, the role of mergers in fueling SMBH growth, and the role of gas content in potentially triggering AGN activity.
        
        \subsubsection{True merger statistics} \label{true-merger-statistics}
            
            \begin{figure} 
                \includegraphics[width=\linewidth]{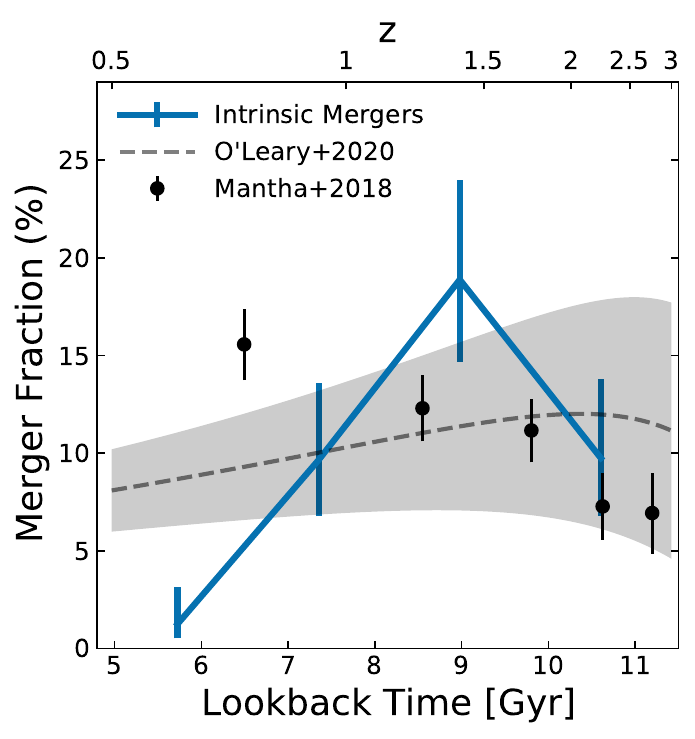}
                \caption{Merger fractions versus lookback time for close pairs identified from the simulation (blue solid). The merger fraction peaks at $19\%$ around $z\sim 1.5$, but plummets to $1\%$ around $z\sim 0.5$. Simulation close pairs exhibit similar merger fractions to those from the \textsc{EMERGE} empirical model \citep{OLeary2021} and to those estimated from CANDELS observations \citep{Mantha2018}, at all times except $z\sim 0.5$.}
                \label{mergerfrac-intrinsic}
            \end{figure}
        
            In Figure \ref{mergerfrac-intrinsic} we compare the time evolution of merger fraction among our $29$ halos. We compare with \citet{Mantha2018} companion fractions of massive galaxies ($M_{\rm star} > 2 \times 10^{10} M_\odot$) with major companions ($>$1:4) out to a projected $50$ kpc, obtained from all five CANDELS fields with corrections for projection effects. We also compare with predicted companion fractions of massive central galaxies ($M_{\rm star} > 10^{10.3} M_\odot$) with major ($>$1:4) stellar companions out to a projected $50$ kpc from the empirical model of \citet{OLeary2021}. Following \citet{Cameron2011}, we calculate the $95$\% binomial confidence interval given the number of mergers identified by each definition, in each time bin.
            
            
            Intrinsic merger fractions are similar to merger fractions from both \citet{Mantha2018} and \citet{OLeary2021} at all but the lowest redshift bin, where merger fractions are lower in the simulation. Close pairs are most common between $z=1-1.5$. It is worth noting that, although our merger fractions are similar to literature values, 1) our sample of halos is not cosmologically representative, and 2) the merger fractions obtained from the simulations are sensitive to the choice of merging timescale chosen. Timescales longer than $\pm300$ Myr will yield more mergers by definition, and hence increase the merger fraction at all times.
        
        \subsubsection{Impact of mergers on SMBH growth}
            \begin{figure} 
                \includegraphics[width=\linewidth]{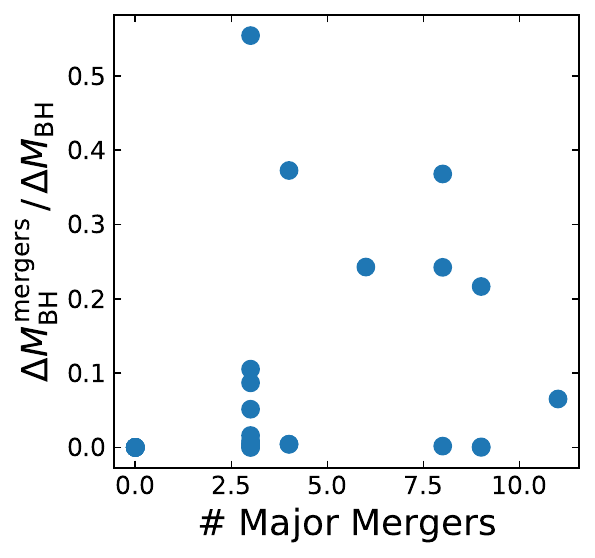}
                \caption{Fraction of $0.5 < z < 3$ SMBH mass growth that occurs during periods of host galaxy mergers, as a function of the number of snapshots containing major merger events. Only one SMBH has the majority of its growth occur during host galaxy mergers. The majority of galaxies grow their central SMBHs little during periods of merger activity. The time a halo spends in mergers does not correlate clearly with how much mass accretes onto SMBHs.}
                \label{mbh-frac-time-frac}
            \end{figure}
    
            An additional goal of our analysis is understanding how important mergers are to the total growth of SMBHs. By tracing the growth histories of the central SMBHs and identifying times when the host galaxy has undergone a merger, we can determine the fraction of total SMBH growth attributable to merger activity.
            
            More time spent in a merging state does not guarantee increased accretion onto SMBHs. Figure \ref{mbh-frac-time-frac} shows the fraction of SMBH mass growth between $0.5 < z < 3$ that occurred during periods of major mergers, as a function of total number of snapshots containing a major merger. Here, $\Delta M_{\rm BH}^{\rm  mergers}$ is the mass grown by the central SMBH during major mergers between $0.5 < z < 3$, while $\Delta M_{\rm BH}^{\rm tot}$ is the total SMBH mass growth between $0.5 < z < 3$. The number of major mergers a galaxy undergoes does not correlate with increased growth onto the central SMBH. Further, only one SMBH had the majority of its growth occur during periods of major merger activity, yet the host galaxy only underwent one major merger event.
            
        \subsubsection{Role of gas content}
        
            \begin{figure} 
                \includegraphics[width=\linewidth]{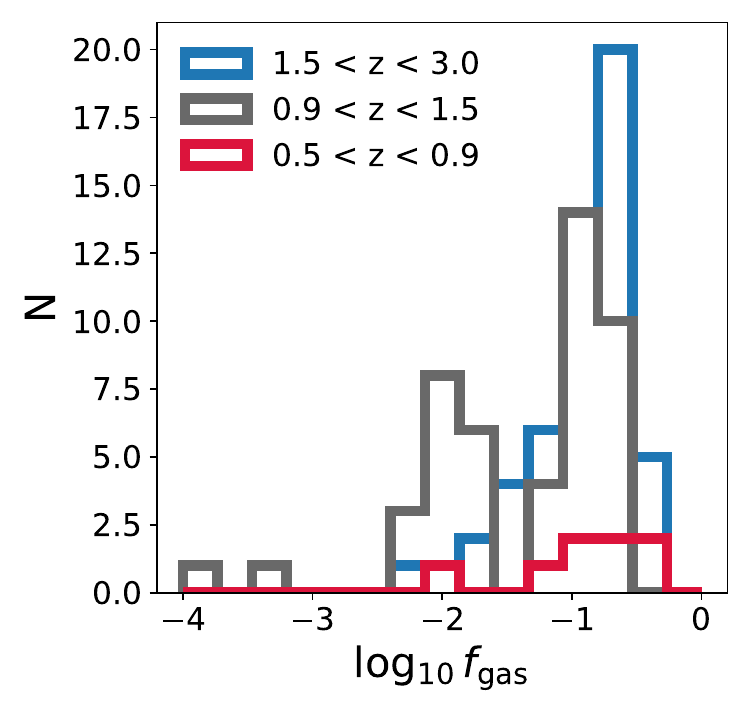}
                \caption{Number of major merging snapshots with a given central galaxy gas fraction, ordered from high redshift (blue) to low redshift (red). Redshift bins are chosen to contain equal numbers of simulation timesteps. Major mergers are overall most common in the between $0.9 < z < 1.5$, but the majority of gas-rich $\left(f_{\rm gas} > 0.1\right)$ mergers occur between $1.5 < z < 3$. At low redshifts $0.5 < z < 0.9$, $6$ of the $8$ merging snapshots contain gas-rich central galaxies.}
                \label{gasrich-gals}
            \end{figure}
            
            \begin{figure*} 
                \includegraphics[width=\linewidth]{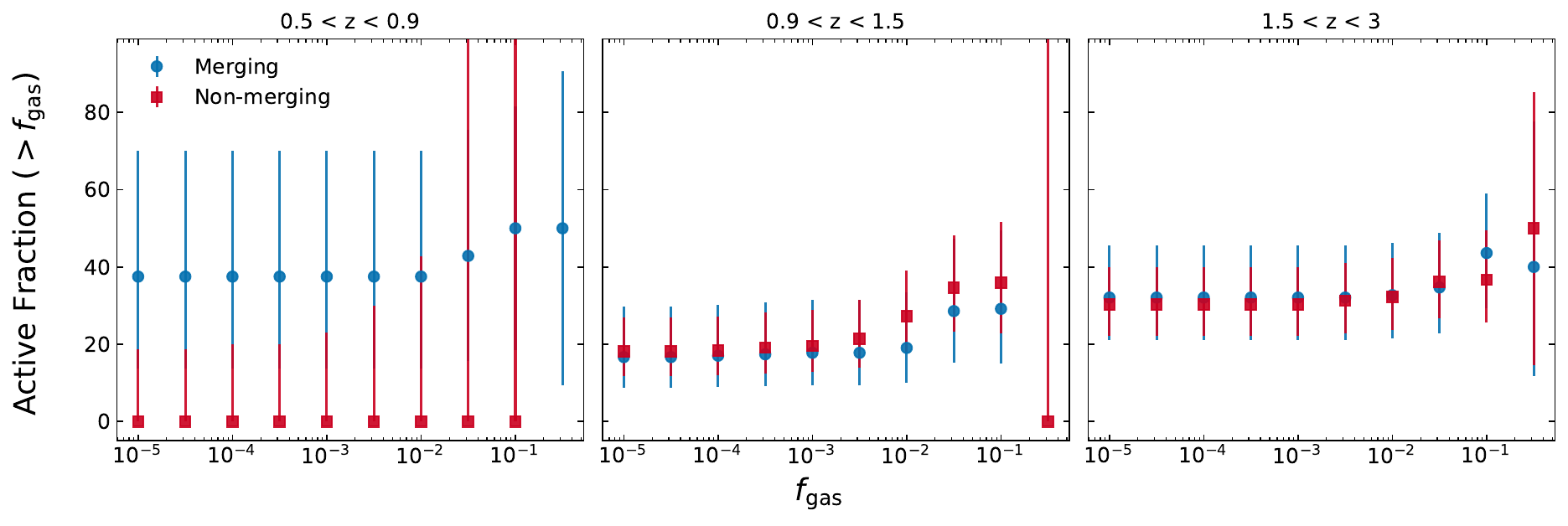}
                \caption{AGN fraction among merging (blue circles) versus non-merging (red squares) galaxies above a given gas fraction threshold, binned by redshift. Redshift bins are chosen as in Figure \ref{gasrich-gals}. We define AGN by a cut on bolometric luminosity, $L_{\rm bol} \geq 10^{43}$ \ergs. AGN fractions are identical between merging and non-merging galaxies above $z > 0.9$. Between $0.5 < z < 0.9$, the few merging galaxies tend to be gas-rich, $3$ of which also host an AGN.}
                \label{agnfrac-gas}
            \end{figure*}
            
            \begin{figure*} 
                \includegraphics[width=\linewidth]{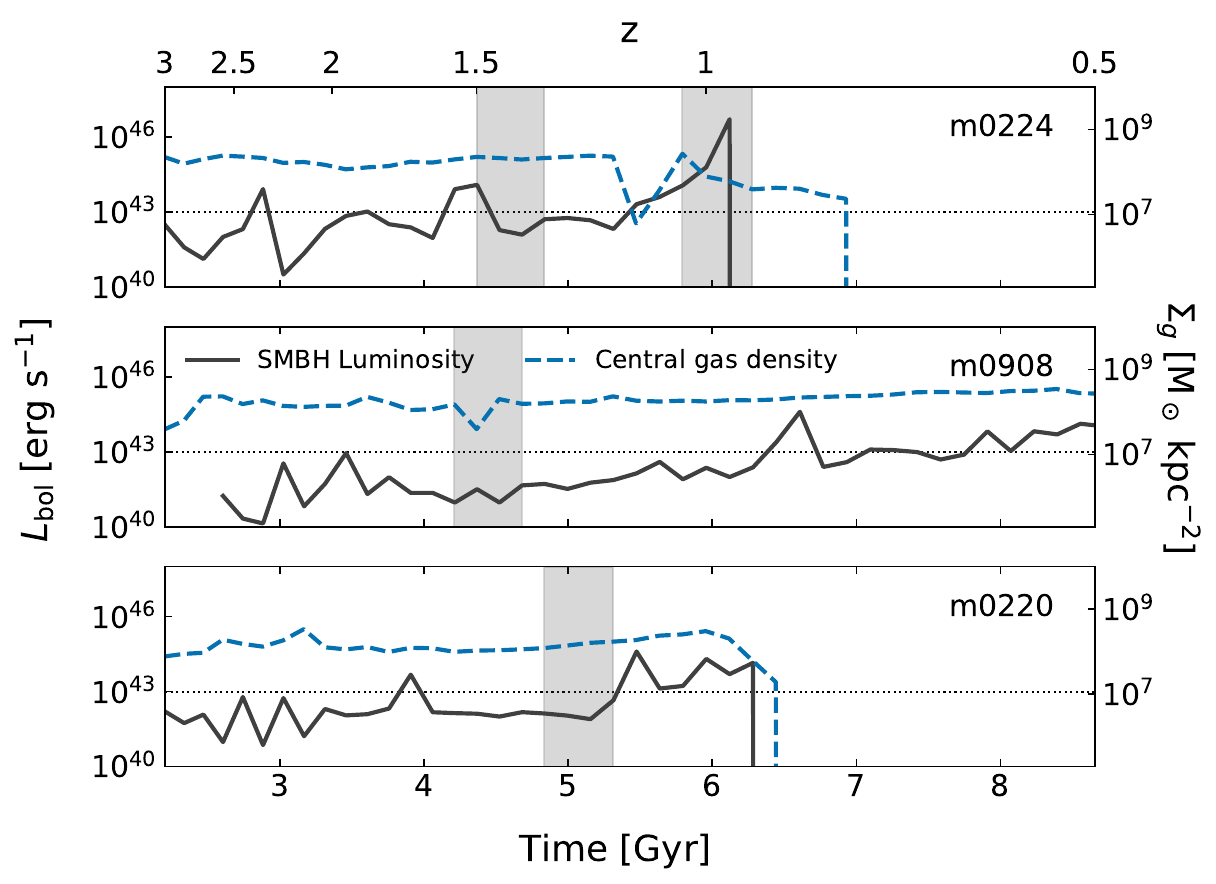}
                \caption{Central SMBH accretion (black solid) and central gas density within $1$ kpc (blue dashed) across cosmic time. We choose $3$ gas rich ($f_{\rm gas} > 0.1$) halos that undergo major mergers. We mark the periods of merger activity with grey shaded regions. The cutoff for activity is marked by a horizontal dotted line. Halo \textit{m0224} appears to have strong AGN activity in response to a merger, while halo \textit{m0908} has no response. Halo \textit{m0220} has an increase in AGN activity during/after the event, with SMBH shutoff $1$ Gyr after the event.}
                \label{sigma-lbol}
            \end{figure*}
    
            We now explore the role of gas content in the ignition of AGN via galaxy interactions. If mergers occur when there is little gas in the central galaxy, it is expected that they would not trigger significant BH growth or AGN activity. On the other hand, gas-rich mergers may be capable of efficiently funneling gas onto the central SMBH \citep[e.g,][]{Blumenthal2018}. 
            
            Figure \ref{gasrich-gals} illustrates the gas fraction within the central galaxy for every snapshot containing a major merger, in bins of redshift between $0.5 < z < 3$. We define the gas fraction by the fraction of all baryons contained in gas:
            \begin{equation}
                f_{\rm gas} = \frac{M_{\rm gas}}{M_{\rm gas} + M_{\rm star}},
            \end{equation}
            where $M_{\rm gas}$ is the gas mass within the central galaxy. Each redshift bin is chosen to contain equal numbers of simulation timesteps between $0.5 < z < 3$. 
            
            Major mergers are most gas-poor during the period of greatest merger activity, between $0.9 < z < 1.5$. Although our galaxies on average become less gas-rich with time, the few major mergers occurring between $0.5 < z < 0.9$ are surprisingly gas-rich $\left(f_{\rm gas} > 0.1\right)$. Hence if gas content plays an important role in triggering AGN, it is either the abundant gas-rich mergers between $1.5 < z < 3$, or the few gas-rich mergers between $0.5 < z < 0.9$, that would be most likely to trigger AGN.
            
           Next we show how the fraction of active galaxies varies with gas fraction. Figure \ref{agnfrac-gas} shows the AGN fraction for central galaxies more gas-rich than a given $f_{\rm gas}$ threshold, binned by redshift. We define AGN by a cut on bolometric luminosity, $L_{\rm bol} \geq 10^{43}$ \ergs. For each merger, we select two non-merging control galaxies. The fraction of AGN in merging systems above $z > 0.9$ is consistent with that of non-merging systems, regardless of gas fraction. In the lowest redshift bin there are only $3$ total AGN, all of which are found in the $8$ total merging systems.
            
            Finally, we choose three central galaxies that were gas rich at $z\sim3$ and track both their central gas density and the bolometric luminosity of the central SMBH. We loosely define gas-rich by $f_{\rm gas} > 0.1$. We also define the central gas density within the central $1$ kpc of the galaxy:
            
            \begin{equation}
                \Sigma_{\rm gas} = \frac{M_{\rm gas}\left(< 1 {\rm kpc}\right)}{4 \pi \left(1 {\rm kpc}\right)^2}.
            \end{equation}
            If mergers are capable of triggering AGN, we expect to see an influx of gas flowing into the central regions following a merger, and hence an increase in central gas density.
    
            Figure \ref{sigma-lbol} shows the evolution of total bolometric luminosity compared with the evolution of the central gas density, with time. We also highlight periods where the central galaxy is in a major merging state.
            
            Halo \textit{m0224} undergoes two major mergers around $z\sim1.5$ and $z\sim0.9$. AGN activity appears to increase immediately preceding both the first and second merger, but only spike sharply in the second. Following a peak in AGN activity during the second merger, it steeply drops, followed by an eventual drop in the central gas content, presumably as gas is evacuated from the central regions. This halo's growth history shows that at least some mergers do appear to be connected with enhanced accretion onto the SMBH, leading to observable levels of AGN activity. 
            
            Halo \textit{m0908} instead undergoes a major merger without significant AGN activity and only a slight variation in central gas density. Instead, AGN activity slowly increases over time as the SMBH grows. The SMBH grows via slow gas accretion from a central gas supply that is constantly replenished by the rest of the galaxy.
            
            Finally, halo \textit{m0220} has one major merger which appears to affect AGN activity in the later stages of the event, as well as after the event. Distinct from the other two halos, AGN activity increases slightly during the merger, then quickly increases a short time after the merger completes. Gas is consumed or ejected until the AGN shuts off $1$ Gyr later, likely from lack of fuel.
            
            Neither halo \textit{m0908} nor halo \textit{m0220} are consistent with gas-rich mergers as the primary driver of AGN activity. This set of halos illustrates that major mergers may sometimes drive immediate episodes of significant black hole accretion, but not all AGN activity is connected with mergers and not all mergers are connected with immediately enhanced AGN activity. Further, it appears that even when galaxies are gas rich, mergers are not guaranteed to trigger significant accretion onto the SMBH. These halos illustrate there are likely additional criteria on top of gas richness that are required in order to trigger AGN activity.
    
    \subsection{Comparison of Intrinsic and Classified Mergers} \label{intrinsic-vs-synthetic}
        
        With intrinsic merger statistics in hand, we now explore the characteristics of our merger classifier applied to the synthetic images. We examine the morphologies exhibited by the central galaxy in each synthetic image, then search for any connection between major mergers identified through both the classifier and intrinsic AGN activity.
        
        When estimating merger fractions derived from the synthetic images, we apply an additional correction for the known incompleteness of the automated classifier, in order to form more equal comparisons with intrinsic merger fractions. Following \citet{Snyder2019}, we define the merger fraction for classified mergers as:
        
        \begin{align}
            f_{\rm merger} = \frac{N_{\rm RF}}{N}\frac{PPV}{TPR},
        \end{align}
        for $N_{\rm RF}$ number of mergers identified by the classifier and $N$ total images. In astronomy terms, $PPV$ is the classifier purity while $TPR$ is the completeness. Multiplying merger fractions by the purity helps account for non-mergers falsely classified as mergers. Dividing by the completeness helps account for true mergers missed by the classifier. The trained classifier yields a correction $PPV / TPR = 0.57$.
        
        \subsubsection{Merger fraction is rarely enhanced among AGN}
        
            \begin{figure*} 
                \includegraphics[width=\linewidth]{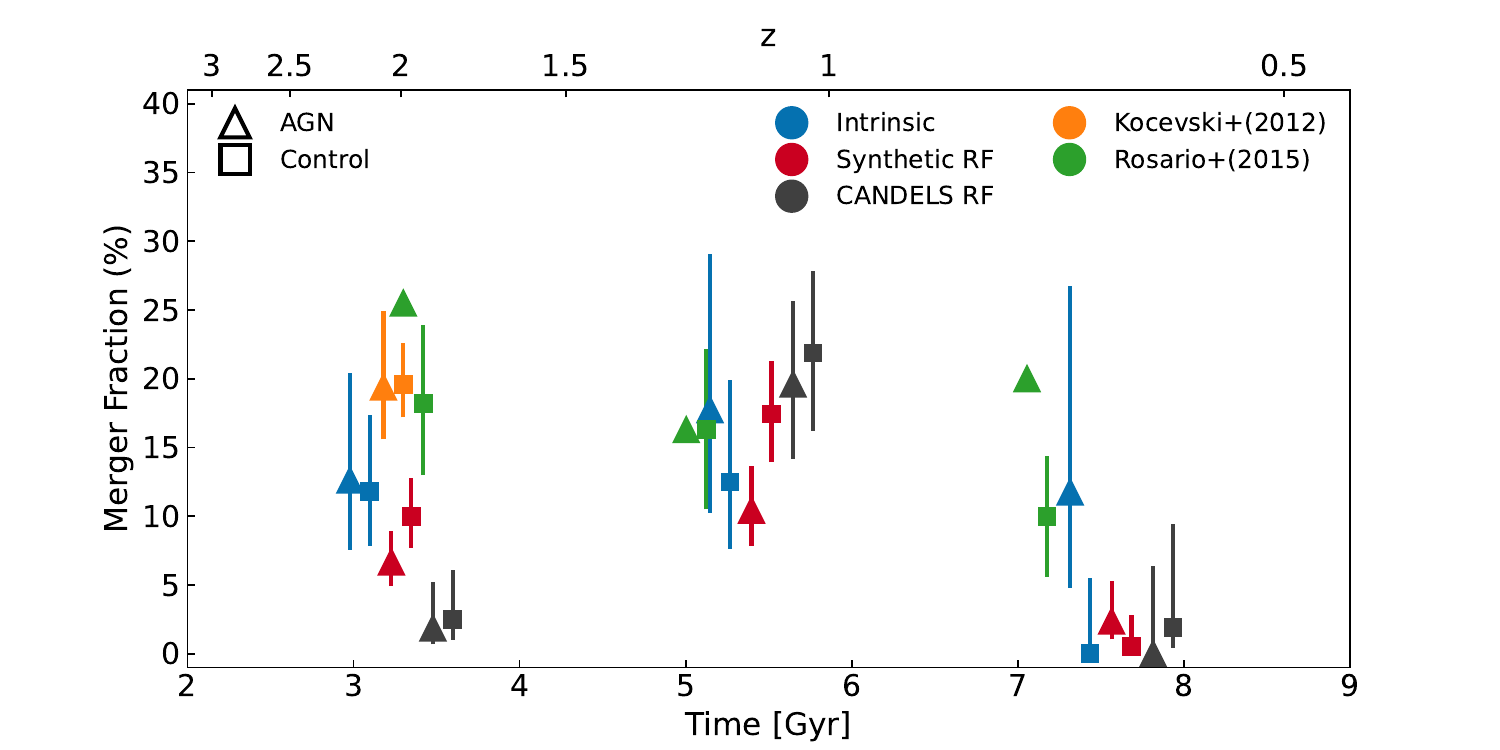}
                \caption{Merger fractions from intrinsic mergers (blue), and RF classifications applied to the synthetic images (red) and CANDELS (black), as a function of lookback time. We distinguish between AGN hosts (triangles) and their inactive, mass-matched control counterparts (squares). We include merger fractions for visually identified disturbed or interacting galaxies in CANDELS \citep{Kocevski2012,Rosario2015}. Intrinsic merger fractions are comparable to observations from \citet{Rosario2015} at all redshifts. The intrinsic and synthetic merger fractions exhibit an excess in AGN at low redshifts, similar to \citet{Rosario2015}.}
                \label{mergerfrac-CANDELS}
            \end{figure*}
        
            \begin{figure} 
                \includegraphics[width=\linewidth]{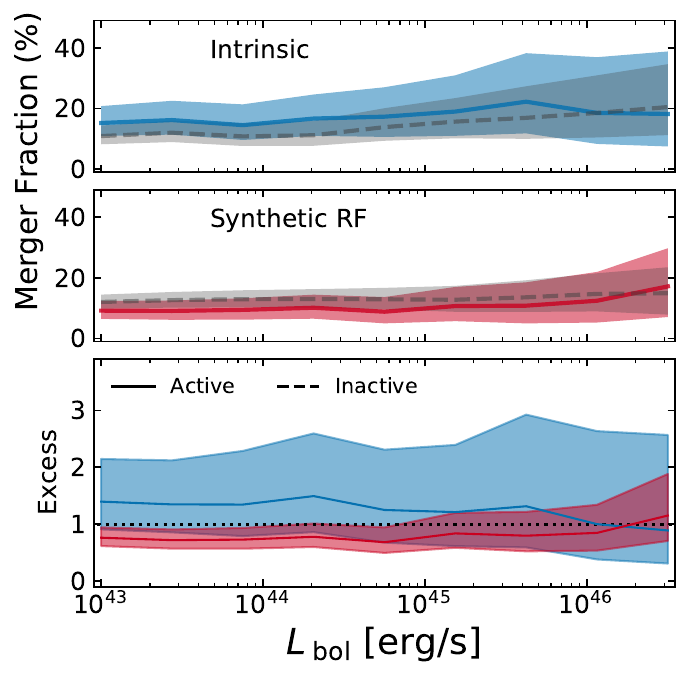}
                \caption{Comparison of merger fractions for AGN (solid) versus non-AGN control (grey, dashed) galaxies, as a function of increasing AGN luminosity threshold. We show merger fractions for both intrinsic (blue, top panel) and RF classified (red, middle panel) mergers. The bottom panel shows the fractional residual between AGN and non-AGN merger fractions, with the $1-1$ relation indicated with a dotted black line. Shaded regions indicate $95$\% confidence intervals. Both intrinsic mergers and classified mergers exhibit no significant difference in merger fractions between AGN and non-AGN, even for the most luminous AGN in our sample ($L_{\rm bol} \sim 10^{46.5}$ \ergs).}
                \label{mergerfrac-Lbol}
            \end{figure}
            
            First, we examine how merger fractions differ across time between AGN and non-AGN. Figure \ref{mergerfrac-CANDELS} compares intrinsic, RF classified synthetic image and CANDELS merger fractions as a function of cosmic time, for galaxies hosting AGN ($L_{\rm bol} > 10^{43}$ \ergs) versus the inactive ($L_{\rm bol} < 10^{43}$ \ergs) control sample. We show the CANDELS merger fraction: companion pair fraction from \citet{Kocevski2012}, who use data from the CANDELS GOODS-S field, and the interaction fractions from \citet{Rosario2015}, who use the combined data of all five CANDELS fields, GOODS-N included. We compute the non-parametric morphology statistics from the CANDELS image and employ them to differentiate between mergers and non-mergers using the Random Forest classifier trained on synthetic images. This process mirrors the application of the RF classifier to the synthetic image for distinguishing between mergers and non-mergers in the observational data. We note that both \citet{Kocevski2012} and \citet{Rosario2015} employed visual classification methods for categorizing mergers.

            In all cases, we found no statistically significant difference in the merger fraction between AGN and the control sample. Intrinsic, RF-classified synthetic images, and RF classified CANDELS images, did not show a significant difference in merger fraction between AGN and non-AGN. However, The few AGN between $0.5 < z < 0.9$ show a slight excess in merger fraction over non-AGN that is not seen at higher redshift. Overall, our intrinsic merger fractions are similar to merger fractions from \citet{Kocevski2012} and \citet{Rosario2015} across time, hovering between $5-15$\%. We find evidence of a slight excess in intrinsic merger fraction at low redshift as seen in \citet{Rosario2015}. This excess is driven by the small number of especially gas-rich AGN found in merging galaxies between $0.5 < z < 0.9$.
            
            Next we explore how merger fraction depends on AGN luminosity between redshifts $0.5 < z < 3$. We make sequentially higher cuts on bolometric luminosity and examine the merger fraction relative to the corresponding inactive control sample. Figure \ref{mergerfrac-Lbol} shows how merger fractions vary with increasing bolometric luminosity thresholds relative to the inactive control samples, in both the intrinsic and classified cases. We also plot the fractional excess merger fraction for AGN versus non-AGN, as a function of the moving luminosity threshold.
        
            Even the most luminous AGN in our sample do not exhibit a significant enhancement of merger fraction relative to the control galaxies. Neither the intrinsic nor the classified merger fractions are significantly different from their corresponding control samples in any luminosity bin.
            
        \subsubsection{AGN fraction is rarely enhanced by mergers}
        
            \begin{figure*} 
                \includegraphics[width=\linewidth]{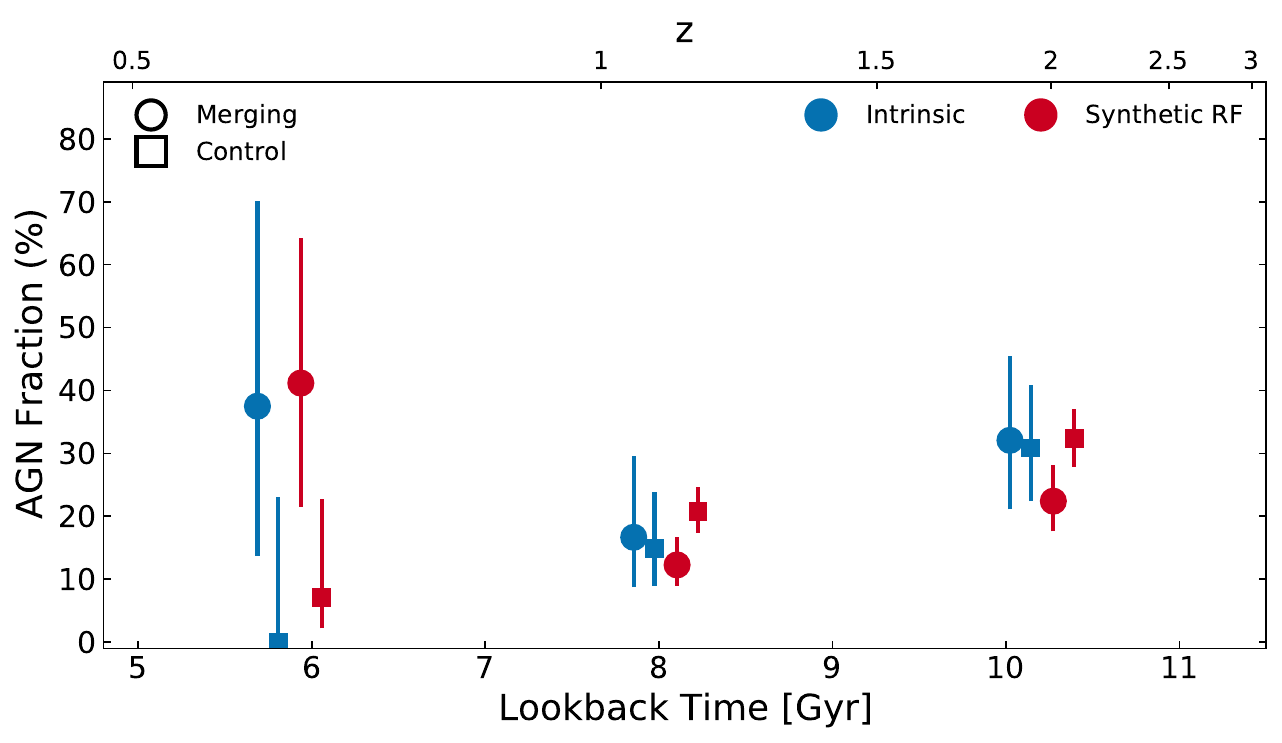}
                \caption{AGN fractions for intrinsic mergers (blue), and RF classifications (red), as a function of lookback time. We distinguish between merging (circles) and non-merging (squares) mass-matched control galaxies. We define AGN by a cut on bolometric luminosity, $L_{\rm bol} \geq 10^{43}$ \ergs. Both merger identifiers show consistent AGN fractions between merging and non-merging systems across all redshifts, except in the lowest redshift bin where mergers exhibit an excess in AGN fraction.}
                \label{agnfrac-mergers}
            \end{figure*}
    
            \begin{figure}
                \includegraphics[width=\linewidth]{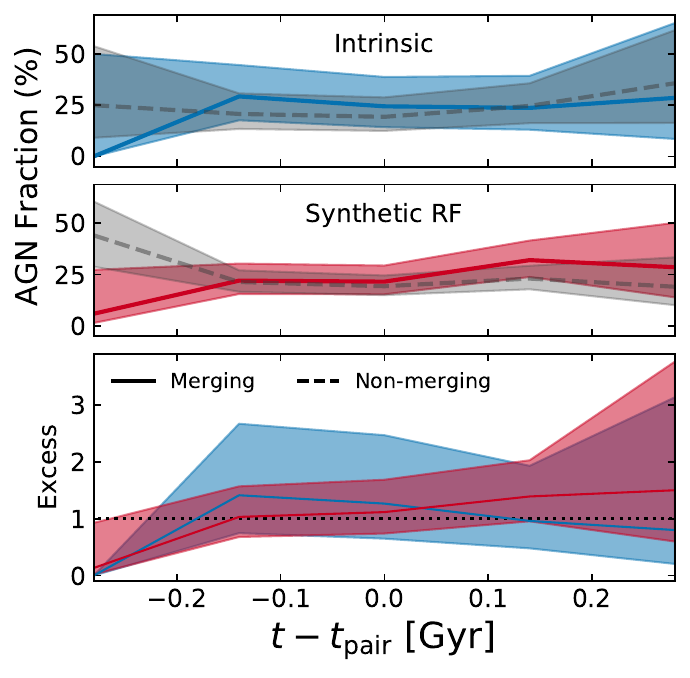}
                \caption{AGN fraction versus relative merger time for merging galaxies (solid) and the corresponding non-merging control galaxies in each bin (grey, dashed). Merger times are relative to the time a close pair was identified, $t_{\rm pair}$, such that positive/negative values correspond to post-/pre-merger states, respectively. We include both intrinsic mergers (blue, top panel) and RF classified mergers (red, middle panel). The bottom panel shows the fractional residual of AGN fractions for both merger identifiers. Regardless of merger identifier, there is no significant excess in AGN fraction at any stage of a merger.}
                \label{agnfrac-mergertime}
            \end{figure}
    
            Previously we asked whether we could detect a difference in the merger fraction of AGN hosts compared with an inactive control sample. We now ask the reverse question: do we see a difference in the AGN fraction of a sample of galaxy mergers, compared to a non-merging control sample?
            
            For each both intrinsic and classified mergers, we select control galaxies which are not experiencing major mergers. We then identify AGN using a bolometric luminosity threshold $L_{\rm bol} > 10^{43}$ \ergs. Figure \ref{agnfrac-mergers} shows the AGN fraction across cosmic time for intrinsic and classified mergers, relative to their corresponding non-merging control samples.
            
            Intrinsic merging systems do not show a significant enhancement of AGN activity relative to non-merging systems. The AGN fraction in merging and non-merging systems does not evolve strongly with time, hovering between $20-30$\% above $z > 0.9$. In the lowest redshift bin, there is an enhancement in AGN fraction, though there few major mergers and few AGN. While we saw an enhancement in merger fraction in low redshift systems, it is unclear how significant the enhancement in AGN fraction is.
            
            Next we track the evolution of AGN fraction throughout each merger period and search for enhancement relative to the control sample. Figure \ref{agnfrac-mergertime} shows AGN fractions for galaxies during each stage of a merger, compared to the AGN fraction of the corresponding control galaxies. We centre the merger times at the snapshot where a close pair was initially identified, $t_{\rm pair}$, such that negative/positive values of $t - t_{\rm pair}$ correspond with pre-/post- merger states, respectively. We also plot the fractional excess in AGN fraction relative to the control at each merger stage, as done previously.
            
            Within $\pm 300$ Myr of pair identification, both intrinsic mergers and those identified from the synthetic images show similar AGN fractions throughout all merger stages, and show no statistically significant difference relative to their corresponding control sets.

\section{Discussion}

    \subsection{Caveats} \label{caveats}

        Our analysis has a few caveats that must be considered, including the impact of torus-scale dust obscuration on identifying AGN, the impact of simulation resolution and sub-grid physics, the subtleties of our methodology, and the impact of our merger detection methods.

        Many studies have found that mid-IR selected AGN show a stronger correlation with mergers than AGN selected at other wavelengths \citep[e.g,][]{Satyapal2014,Ellison2019}. \citet{Donley2018} find that IR-selected, heavily obscured AGN in the CANDELS/COSMOS field are more likely than X-ray-selected AGN to be identified as disturbed, potentially post-merger, systems. \citet{Kocevski2015} find heavily obscured AGN at $z\sim1$ are three times more likely than unobscured AGN to show signatures of galaxy interactions.
        
        Although our RT model includes galaxy-scale dust, it does not include dust on the scale of the accretion structure surrounding the black hole. While we can directly measure intrinsic accretion rates onto the central SMBHs, we do not take into account the torus-scale obscuration in the generation of our synthetic images, or in our AGN luminosity cuts. The inclusion of torus-scale obscuration may be a relevant ingredient in realistically selecting AGN from the simulations, by potentially removing the few merger-induced AGN from detection. Future work may benefit from more detailed modeling of AGN emission \citep[e.g,][]{Done2012} and torus-scale obscuration \citep[such as the models detailed in][]{Nenkova2008,Nenkova2008a}.
        
        Simulation resolution and sub-grid physics could also impact our results. Our choices of SMBH accretion and feedback physics within the simulation directly impact the measured AGN activity, and are themselves dependent on simulation resolution. In order to study the impact of mergers on AGN activity, both AGN scales and cosmological scales must be simulated at high resolution. Simulations run at higher spatial resolution would allow the inclusion of more detailed physics, that occurs well below the resolution limit of our simulations. For example, hyper-zoom simulations that study accretion onto SMBH find that the Bondi-Hoyle approximation used in our simulation suite may be quite inaccurate \citep{Angles-Alcazar2021}. \citet{Zanisi2021} find that higher resolution cosmological hydrodynamic simulations are better able to capture the small-scale galaxy morphologies, which may be tied to AGN response to mergers. Higher resolutions and more realistic modelling may find a stronger and more broad correlation between merger events and induced AGN activity.

        Due to the nature of high-resolution zoom-in simulations, we do not have a cosmologically representative sample relative to larger, uniform cosmological simulations and observational surveys. Although we can increase our image sample size by observing each galaxy from multiple orientations, we likely suffer from systematics. For example, our simulation suite does not include the very massive systems that likely give rise to the most luminous AGN and quasars at high redshifts, and may not provide good statistical representation of rare major mergers. Indeed, it is the most massive and luminous galaxies that have been observed to have the strongest connection between mergers and AGN \citep[e.g,][]{Goulding2018}. With only $29$ halos, we are likely not sampling the full diversity of halo merger histories.
        
       Many observations rely on non-parametric morphological parameters to differentiate between mergers and non-mergers. However, our simulations and the analysis of mock images have revealed that these parameters lack effective discrimination between intrinsic mergers and non-mergers when employed individually, as illustrated in \autoref{roc-curve}. When combining all these parameters and employing machine learning techniques, we managed to create a more robust merger classifier than using them individually, but its performance remains suboptimal. Our conclusion is that employing non-parametric measurements for classifying mergers might not be advisable initially. Exploring alternative machine learning methods that directly incorporate mock images could prove to be a more fruitful way to classify merger.
       
       In addition, as seen in Figure \ref{confusion-matrix}, our automated classifier performs well in identifying true mergers at the cost of incorrectly classifying many non-mergers. The high impurity in our classifier means that it is often difficult to disentangle the effects of misclassification from effects of the mock observation process itself. In falsely classifying non-mergers as mergers, true connections between mergers and AGN may be washed out. For example, in Figure \ref{mergerfrac-CANDELS} we found an excess of intrinsic merger fraction with high uncertainty at low redshift. In moving to the classifier merger fractions, the uncertainties are lower but the excess is smaller. It is not clear whether the discrepancy in merger fraction is due the classifier impurity washing out any intrinsic enhancement, or if there is an implicit bias away from mergers in AGN within the methodology itself. Machine learning methods with higher classification precision \citep[e.g,][]{Huertas-Company2019} may find differing results.

        An imbalance in the class of data also greatly affects the performance of RF classifier. To overcome this problem, we perform a random undersampling of the majority class (non-merger), but this sampling is not perfect. There are various ways to reduce the imbalance of data, such as including more snapshots when mergers occur, and we leave this as future work.

        Finally, there are many ways that mergers can be identified both within the simulation and in observations. As discussed in Section \ref{merger-identification}, we choose to identify major mergers within the simulation based on a close pairs method. Similar studies \citep{Rodriguez-Gomez2015,Abruzzo2018} instead generate halo merger trees to identify galaxy mergers and trace progenitor histories. The identified galaxy mergers instead depend on different observing timescales from close pairs. It is not clear how the choice of merger selection method or merger tree generation algorithm affects the end sample of selected mergers, or even if such methods are directly comparable. 

    \subsection{Interpretation and Implications}
    
        Our study has yielded some important insights into the AGN-merger connection and our ability to probe it with observations. It is clear that the simple picture painted by binary merger simulations is not representative of the results of modern cosmological simulations. The reasons for these differences are not yet entirely clear, and will be the subject of future studies, but we can speculate on some of them. The most commonly used initial conditions for binary merger simulations in the past have been 1:1 mergers of gas-rich, thin, rotationally supported (Milky Way like) spiral galaxies \citep[e.g,][]{Springel2005,Hopkins2005,Cox2006,Bournaud2008}. These are somewhat different from the progenitors of most AGN hosts, particularly at high redshift. Based on both simulations \citep{Thomas2005,Somerville2015,Naab2017,Fiacconi2017} and observations \citep{Mullaney2012a,Santini2012,ForsterSchreiber2014,Elmegreen2017,ForsterSchreiber2018,ForsterSchreiber2020}, the progenitors of AGN hosts are more likely to be highly turbulent and thickened star-forming galaxies. Indeed, very gas-rich galaxies are actually thought to be \emph{less} susceptible to strong merger-driven nuclear inflows, as most of the torque that drives gas inwards is caused by the lag between the gaseous and stellar bar \citep{Robertson2006,Hopkins2009a}. Moreover, there have been few studies on the impact of minor mergers on AGN triggering, though studies of nuclear inflows and starbursts indicate that minor mergers would be expected to have a less dramatic impact \citep{Cox2008}. The response of the disk to a perturber is also sensitive to the radial extent and presence of a pre-existing bulge in the primary, and to the satellite orbit \citep{Cox2008}. Finally, idealized binary merger simulations do not include cosmological accretion or accretion of cold gas from a hot halo.
        
        Moving from idealized merger simulations to cosmological simulations does not entirely solve the problem, since modern cosmological simulations also give varying results. \citet{McAlpine2020} find that mergers strongly correlate with an enhancement of AGN fraction within gas-rich $M_{\rm star} \sim 10^{10} M_\odot$ galaxies in the \textsc{EAGLE} cosmological simulation, in tension with our findings of similar mass galaxies. Both this work and \citet{McAlpine2020} find 1) similar merger and AGN fractions across time, and 2) that mergers are not a significant growth mode for the BH, either because accretion enhancement is too little, mergers do not last long enough, or there are too few significant mergers. Using the ``genetic modification" technique on a zoom-in cosmological simulation of similar resolution to our simulations, \citet{Pontzen2017} are able to alter the merger history of a halo with virial mass $M_{\rm vir} \sim 10^{12} M_\odot$, while keeping the final halo mass, large-scale structure, and local environment unchanged. With each iteration of the reference halo, they are able to measure the impact of smaller/larger infalling companions on AGN activity and galaxy quenching. They find that mergers are able to trigger brief periods of SMBH accretion, but not long term growth. On the other hand, \citet{Steinborn2018} find that recent mergers in the Magneticum Pathfinder cosmological simulation can increase the probability of subsequent AGN activity by a factor of three, but overall can only account for the population of very luminous AGN $(L_{\rm bol} > 10^{46}$\ergs$)$ at $z\sim2$.
        
        Ultimately, all simulation results are impacted by the simulation characteristics and implementation of physics. Given the interaction of all of the physical factors with the expected sensitivity to the sub-grid recipes for star formation, stellar feedback, black hole accretion, and to resolution, it is perhaps not surprising that even recent cosmological simulations present a mixed picture of the AGN-merger connection. More detailed studies that look at the effect of mergers in different simulations while controlling for these other variables (progenitor structure and morphology, gas fraction, orbit, etc) will be needed to begin to disentangle these effects.
        
        Another important conclusion of our study is that we do not expect to see significant enhancement of AGN activity in early stage mergers, when galaxies can be identified as close pairs. Any enhancement is expected to manifest in late stage mergers, after the nuclei have coalesced, as seen in \citet{McAlpine2020}. Traditional methods used to identify mergers from images, including those based on non-parametric morphological statistics, are not even very effective at identifying early stage mergers --- identifying a post-merger system is likely much more difficult. However, pixel-based deep learning techniques such as those utilized by \citet{Bottrell2019,Ferreira2020,Bickley2021,Zanisi2021} may be able to identify these late-stage mergers more effectively, which is something that we can test with our synthetic images.
        
        Our results are consistent with the observational CANDELS sample that was the target comparison sample for our study, and somewhat agree with observational studies that have found evidence for significant enhancement in the merger fraction of AGN hosts, and of the AGN fraction of mergers/post-mergers around $z\sim0$ \citep[e.g.][]{Satyapal2014,Ellison2019}. \citet{Rosario2015} suggest that evolving gas fractions may mean mergers are \emph{required} to trigger AGN in the local universe, but are not necessary at high redshifts, in line with what we find in our analysis. We emphasize that in order to assess the consistency between a particular simulation and a particular observational analysis, it is critical to carry out the analysis in a consistent manner on synthetic observations. Such synthetic observations should be created with great care to match the observational characteristics of the sample to be compared with \citep{Bottrell2019,Zanisi2021}.

\section{Conclusion}

    In this paper we explore the potential connection between galaxy mergers, AGN activity, gas content, and overall SMBH growth in a set of cosmological zoom-in simulations between $0.5 < z < 3$. We identify mergers within the simulation by finding close pairs ($<50$ kpc separation) of major $\left(1:4\right)$ stellar companions. We run the dust radiative transfer code \textsc{Powderday} in post-processing to generate mock observations which emulate the characteristics of the CANDELS survey. We then train a random forest classifier to identify mergers based on non-parametric morphological statistics measured from the synthetic images. This method allows us to make reasonable comparisons with observational statistics from CANDELS.
    
    By comparing mergers identified directly from the simulation to those identified through the classifier, we study how AGN activity may correlate with merger events. With carefully selected mass-matched control samples, we compare how merger fractions vary between AGN and inactive galaxies, as well as how AGN fractions vary between merging and non-merging systems. We find that:

    \begin{enumerate}
        \item Most SMBHs in the simulations do not grow significantly during periods of merger activity. The amount of time that a galaxy spends in a merging state has no clear bearing on how much a SMBH grows.
        \item Gas-rich mergers are most common at high redshift $\left(1.5 < z < 3\right)$, and mergers are overall most common at intermediate redshift $\left(0.9 < z < 1.5\right)$, yet neither redshift range show evidence of merger-induced AGN activity. 
        \item Mergers occur among AGN with the same probability as among non-AGN, except at low redshift $\left(0.5 < z < 0.9\right)$ where the very few gas-rich mergers tend to host AGN. This result is in-line with \citet{Rosario2015} who find that evolving global gas fractions mean that mergers are required to trigger AGN in the local universe. Moreover, highly luminous AGN ($L_{\rm bol} = 10^{43-46.5}$ \ergs) are no more likely than non-AGN to undergo mergers.
        \item AGN can be found with equal probability in merging as non-merging systems, except at low redshift $\left(z\sim0.5\right)$ where AGN are slightly more common in merging systems. On average, AGN activity is not enhanced at any particular phase of a major merger.
        \item The predictions of our state-of-the-art zoom cosmological simulations for the AGN-merger connection are consistent with the CANDELS observational sample, when the analysis is carried out in a reasonably comparable manner.
        
    \end{enumerate}

    Together this evidence suggests that, while gas-rich mergers may sometimes trigger observable AGN activity at low redshift (and indeed may be a requirement to observe AGN at low redshift), there are probably other mechanisms that play a significant role in fueling moderate luminosity AGN and driving the growth of SMBHs. We plan to further investigate the role of torus-scale obscuration on AGN selection, and hence on the detection of an AGN-merger connection in future work. 

\section{Acknowledgements}
    We thank the anonymous referee for providing many valuable comments and suggestions.
We thank Alyson Brooks, Ariyeh Maller, and Christopher Hayward for helpful conversations.
    This work was supported by the 2023 Research Fund of the University of Seoul for E.~Choi. Also, this work is supported by NASA grant HST-AR-14287.003 for R.~Sharma.
R.~Somerville gratefully acknowledges support from the Simons Foundation. TN acknowledges support from the Deutsche Forschungsgemeinschaft (DFG, German Research Foundation) under Germany’s Excellence Strategy - EXC-2094 - 390783311 from the DFG Cluster of Excellence ``ORIGINS”. We acknowledge the Office of Advanced Research Computing (OARC) at Rutgers, The State University of New Jersey for providing access to the Amarel cluster and associated research computing resources that have contributed to the results reported here.
        
    This work makes use of the following software packages: \textsc{Powderday} \citep{Narayanan2021}, \textsc{yt} \citep{Turk2011}, \textsc{Hyperion} \citep{Robitaille2011}, \textsc{FSPS} \citep{Conroy2010}, \textsc{statmorph} \citep{Rodriguez-Gomez2019}, \textsc{RealSim} \citep{Bottrell2019}, \textsc{Scikit-learn} \citep{Pedregosa2011}, \textsc{corner.py} \citep{Foreman-Mackey2016}, and \textsc{Astropy} \citep{AstropyCollaboration2013, AstropyCollaboration2018}. Powderday was conceived of at the Aspen Center for Physics. The Aspen Center for Physics is supported by National Science Foundation grant PHY-2210452.

\section{Data Availability}
The data underlying this article will be shared on reasonable request to the corresponding author.



\bibliographystyle{mnras}
\bibliography{library} 



\bsp	
\label{lastpage}
\end{document}